\documentclass[apj]{emulateapj}

\usepackage{graphicx,natbib,color}

\shorttitle{Electron transport in solar wind density turbulence }
\shortauthors{H. A. S. Reid and E.P. Kontar}

\begin{document}

\title{Solar wind density turbulence and solar flare electron transport from the Sun to the Earth}

\author{Hamish A. S. Reid and Eduard P. Kontar}
\affil{Department of Physics and Astronomy,
University of Glasgow, G12 8QQ, United Kingdom}
\email{hamish@astro.gla.ac.uk, eduard@astro.gla.ac.uk}

\begin{abstract}

Solar flare accelerated electron beams propagating away from the Sun can interact with the turbulent interplanetary media, producing plasma waves and type III radio emission. These electron beams are detected near the Earth with a double power-law energy spectrum. We simulate electron beam propagation from the Sun to the Earth in the weak turbulent regime taking into account the self-consistent generation of plasma waves and subsequent wave interaction with density fluctuations from low frequency MHD turbulence. The rate at which plasma waves are induced by an unstable electron beam is reduced by background density fluctuations, most acutely when fluctuations have large amplitudes or small wavelengths. This suppression of plasma waves alters the wave distrubtion which changes the electron beam transport. Assuming a 5/3 Kolmogorov-type power density spectrum of fluctuations often observed near the Earth, we investigate the corresponding energy spectrum of the electron beam after it has propagated 1~AU. We find a direct correlation between the spectrum of the double power-law below the break energy and the turbulent intensity of the background plasma. For an initial spectral index of 3.5, we find a range of spectra below the break energy between 1.6-2.1, with higher levels of turbulence corresponding to higher spectral indices.

\end{abstract}

\keywords{Sun:flares - Sun: X-rays, gamma rays - Sun:activity -Sun: particle emission}

\section{Introduction}

Solar flare impulsive electron events present an alternative to the more traditional hard X-ray diagnostics of poorly understood acceleration and transport of solar energetic electrons.  While hard X-ray observations provide insight into energetic electrons in the lower dense solar atmosphere \citep[e.g.][]{Arnoldy_etal1968,DennisSchwartz1989,BrownKontar05}, impulsive solar electron events \citep[e.g.][]{Lin1985,Krucker_etal07} provide crucial information about escaping electrons from the acceleration region.  Because of the rather limited spatial resolution of past and current hard X-ray observations, even the spatially resolved hard X-ray spectrum of energetic electrons with RHESSI \citep[e.g.][]{Emslie_etal2003} is a convolution of transport effects and possibly electron acceleration \citep{Brown_etal2009}. Hence, de-convolution of the electron accelerated spectrum and accelerator properties from hard X-rays \citep{Brown_etal2006} is a non-trivial task using both forward-modelling \citep{Holman_etal2003,Kasparova_etal2005} and model-independent techniques \citep{Piana_etal2003,Kontar_etal2005}.  Solar electron impulsive events propagate outward through the almost collisionless plasma of the solar corona and solar wind \citep{Lin1985}.  Even with this collisionless regime the energetic electrons can interact with plasma via generation and absorption of electrostatic plasma waves. In the standard scenario \citep{GinzburgZhelezniakov1958}, the non-linear interaction of beam-driven plasma waves leads to the appearance of rather strong radio emission - type III solar/interplanetary radio bursts. The observations of type III solar bursts and energetic particles \citep{Linetal1981,Ergun_etal98,Gosling_etal2003,Krucker_etal07} as well as theoretical \citep{ZheleznyakovZaitsev1970,Zaitsev_etal72,Melnik1995} and numerical investigations \citep{MagelssenSmith1977,Grognard1982,Kontar_etal1998,Yoon_etal2000,Kontar2001_a,Li_etal2006a,Ledenev04,Krasnoselskikh_etal2007} provide strong support to the standard type III model.

The plasma of the solar corona and the solar wind is a non-uniform turbulent medium with density perturbations at various length scales. The structure of the solar wind density fluctuations have been analysed using scintillations of small-size radio sources \citep[e.g.][]{Hollweg1970,Young1971}. In-situ measurements have also been used to determine the density spectrum near the Earth and between $0.3$ and $1$~AU with {\it Helios} \citep{MarschTu1990}. While the detailed structure of the density turbulence in the inner heliosphere is not well established, the density fluctuation spectrum near the Earth seems close to a power-law spectrum with spectral index about $5/3$, similar to earlier observations. It has been recognized \citep{Ryutov1969,KarpmanIstomin1974} that beam-driven Langmuir waves can be effectively altered by even weak density gradients. Therefore the effect of density fluctuations on beam-driven plasma waves responsible for type III radio bursts has been considered both numerically and analytically \citep{Melroseetal1986,Robinsonetal1992,Kontar2001_solphys}. Density fluctuations are believed to suppress plasma wave growth \citep{SmithSime1979,Muschiettietal1985} and be responsible for the clumpy plasma wave distribution observed in-situ near the Earth \citep{Gurnettetal1978,Linetal1981}. The fluctuations, whilst changing the distribution of plasma waves significantly, have a rather weak modulation effect on the instantaneous distribution of electrons \citep{Kontar2001_a}. Recently, \citet{KontarReid2009} have shown that the electron beam plasma interaction via Langmuir waves in the non-uniform solar corona leads to the appearance of a break energy in the observed spectrum at the Earth and can explain the observed apparent early injection of low-energy electrons. However, the net effect of density fluctuations in the solar wind  on the electron spectrum detected near 1~AU has  not been addressed before.  

In this paper, we investigate the effects of background plasma density fluctuation on the generation and absorption of plasma waves from a high energy solar electron beam travelling from the Sun to the Earth. We demonstrate the dependence of plasma waves on the level of density fluctuations, with high levels damping plasma waves too much to be in accordance with detected type III radio emission. We also show how the level of density fluctuations has a direct effect on the spectral characteristics of the electron beam near the Earth.

\section{Electron Beam Transport Model}

There is a variety of different processes which affect the propagation of high energy electrons from the solar corona through the heliosphere \citep[see][as a review]{Melrose1990}. This work focusses on the role of electron beam-driven electrostatic turbulence in the propagation and spectral evolution of energetic particles. The electrostatic turbulence plays the dominant role for deca-keV electrons. The solar magnetic field expanding into the heliosphere quickly decreases with distance and provides adiabatic focussing for energetic electrons which ensures one dimensional (along expanding magnetic field lines) electron transport. To describe self-consistently resonant interaction of the electron distribution function $f(v,r,t)$ (the number density of energetic electrons is $n_b=\int f dv$) and the spectral energy density of electron plasma waves $W(v,r,t)$ (the energy density of plasma waves is $\int W dk$ ergs cm$^{-3}$) in the radially expanding magnetic field of the heliosphere, one can use the following equations of weak turbulence theory
\begin{eqnarray}
\frac{\partial f}{\partial t} + \frac{v}{(r+r_0)^2}\frac{\partial}{\partial r}(r + r_0)^2f =
\frac{4\pi ^2e^2}{m^2}\frac{\partial }{\partial v}\frac{W}{v}\frac{\partial f}{\partial v}  \;\cr +\frac{4\pi n_e e^4}{m_e^2}\ln\Lambda\frac{\partial}{\partial v}\frac{f}{v^2}
\label{eqk1}
\end{eqnarray}
\begin{eqnarray}
\frac{\partial W}{\partial t} + \frac{\partial \omega_L}{\partial k}\frac{\partial W}{\partial r}
-\frac{\partial \omega _{pe}}{\partial r}\frac{\partial W}{\partial k}
= \frac{\pi \omega_{pe}}{n_e}v^2W\frac{\partial f}{\partial v} \;\;\;\;\;\;\;\cr - (\gamma_{c} +\gamma_L )W + e^2\omega_{pe}(r) v f \ln{\frac{v}{v_{Te}}}.
\label{eqk2}
\end{eqnarray}
The first terms at the right hand sides of Equations (\ref{eqk1},\ref{eqk2}) describe the resonant interaction, $\omega_{pe} = kv$ of electrons and plasma waves first derived by \citet{Drummond_Pines1962,Vedenov_etal1962}. The dispersion relation of plasma waves is  $\omega_{L}(k)=\omega_{pe}+3v_{Te}^2k^2/(2\omega_{pe})$, so the group velocity of plasma waves is $\partial \omega_L/\partial k = 3v_{Te}^2/v$ in Equation (\ref{eqk2}) where $v_{Te}=\sqrt{k_BT_e/m_e}$.  Following \citet{ZheleznyakovZaitsev1970,TakakuraShibahashi1976} we include collisional losses both for electrons and Langmuir waves. The last term of Equation (\ref{eqk1}) accounts for electron collisional Coulomb losses in fully ionized hydrogen plasma \citep[e.g.][]{Emslie1978}, $\gamma_{c} = {\pi n_e e^4}\ln\Lambda/(m_e^2 v_{Te}^3)$ is the collisional damping rate of Langmuir waves, and $\gamma_L=\sqrt{2\pi}\omega_{pe}\left(v/v_{Te}\right)^3\exp\left(-\frac{v^2}{v_{Te}^2}\right)$ is the Landau damping of Langmuir waves by background plasma. The last term in Equation (\ref{eqk2}) is the spontaneous wave generation, which is similar to \citet{ZheleznyakovZaitsev1970,TakakuraShibahashi1976,Hannah_etal2009} but different from the terms used in \citet{Li_etal2006b}. We note that for large velocities ($v \gtrsim v_{Te}\sqrt{2\ln\Lambda}$) the energy loss of an electron to spontaneously generate Langmuir waves adopted by \citet{Li_etal2006b} is greater than the electron collisional Coulomb losses in fully ionized hydrogen plasma (last term of Equation \ref{eqk1}).

The second term on the left hand side of Equation (\ref{eqk1}) models magnetic field expansion from the solar corona into interplanetary space and the `origin' of the field cone $r_0=3\times10^9$~cm is chosen to have the cone expansion of $33.6^o$.  The heliospheric expansion conserves the total number of electrons such that for scatter-free propagation, $\int(r+r_0)^2 n(r) dr = const$.

The effect of the background electron density gradient on plasma waves is governed by the last term on the left hand side of Equation \ref{eqk2}. Similarly to \citet{Kontar2001_solphys}  we define the characteristic scale of plasma inhomogeneity, $L=\omega_{pe}(\partial \omega_{pe}/\partial r)^{-1} = 2n_{e}(\partial n_{e}/\partial r)^{-1}$. This values has to be larger than the wavelength of any plasma waves considered to remain within the Westzel-Kramers-Brillouin (WKB) approximation of geometrical optics. 

\subsection{Initial conditions}

The electron distribution function is modelled using an instantaneous electron injection which is Gaussian in space with a characteristic size $d$. This electron distribution has a power-law spectrum in velocity, and hence in energy, as often observed in solar flares \citep[e.g.][]{BrownKontar05}.  $f(v,r,t=0)$ takes the form
\begin{equation}
f(v,r,t=0) = \exp\left(-\frac{r^{2}}{d^{2}} \right)\frac{n_{b}(2\delta -1) }{v_{min}}\left(\frac{v_{min}}{v}\right)^{2\delta}.
\label{init_f}
\end{equation}
The electron beam is normalised to the electron number density $n_{b}$. $\delta$ represents the spectral index of the energy power-law and $v_{min}$ represents the minimum velocity used for the electron beam. 

The initial location of an electron beam ($r=0$ in the above equations) for the subsequent simulations is taken at a background plasma frequency of $500$~MHz which corresponds to the height of $3\times 10^{9}$~cm$^{-3}$ above the photosphere.  This is often interpreted as the typical frequency/location for an electron beam acceleration site in the corona \citep{Aschwandenetal1995}. The spectral index $\delta$ was set to 3.5, corresponding to typical spectral indices above the break energy of in-situ measured electron beams at the Earth \citep{Krucker_etal09}.  The beam size was taken to be $d=10^9$~cm.  Electron thermal velocity was taken to be $v_{Te}=5.5\times 10^{8}$~cm/s, which corresponds to Maxwellian plasma with a temperature of 1~MK. The beam velocities will range between $3.6v_{Te}\approx 2\times 10^{9}$~cm/s and $2\times10^{10}$~cm/s. Above the maximum velocity relativistic effects become important.  Plasma waves created near thermal velocity are absorbed by the background Maxwellian through Landau damping so $3.6v_{Te}$ is an acceptable lower limit.  

The initial electron beam density is taken to be $1.1\times 10^5$~cm$^{-3}$ which, together with $\delta=3.5$,  gives the total number of electrons above 50~keV of $1.2(\sqrt{\pi}d)^3\approx 7\times 10^{27}$.  This is a relatively small event in relation to observed number of electrons above 50~keV \citep{Krucker_etal07}.  The instantaneous injection of the electron beam restricts the total injected electrons to small event sizes to keep the flux of electrons around $100$~keV near the Earth in line with typical values observed at 1~AU \citep{Krucker_etal07,Krucker_etal09}.  If we consider similar number densities at the peak of a temporal Gaussian injection of order $10^3$~s, the total number of electrons rises to $10^{31}$ in agreement with observations \citep{Krucker_etal07}.

The initial spectral energy density of the plasma waves is assumed to be at the thermal level
\begin{equation}\label{init_w}
W(v,r,t=0) = \frac{k_BT_e}{4\pi^2}\frac{\omega_{pe}(r)^2}{v^2}\log\left(\frac{v}{v_{Te}}\right)
\end{equation}
where $T$ is the background plasma temperature, $k_B$ is Boltzmann constant and $v_{Te}$ is the background electron thermal velocity. This thermal level is formed setting $dW/dt=0$ for Maxwellian distribution of electrons with temperature $T$ and ignoring electron collisions in Equation (\ref{eqk2}).

\subsection{Heliospheric plasma density}

The background heliospheric plasma is modelled as a continuously decreasing background electron density. The background density model is found using the equations for a stationary spherical symmetric solution for the solar wind \citep{Parker1958} with a constant found by satellite measurements near the Earth's orbit \citep{Mannetal1999}.  See \citet{KontarReid2009} for details.  The model is static in time because the characteristic electron beam velocities are much larger than solar wind speeds.

\section{Electron transport through plasma with decreasing density}

\begin{figure} \center
\includegraphics[width=0.99\columnwidth]{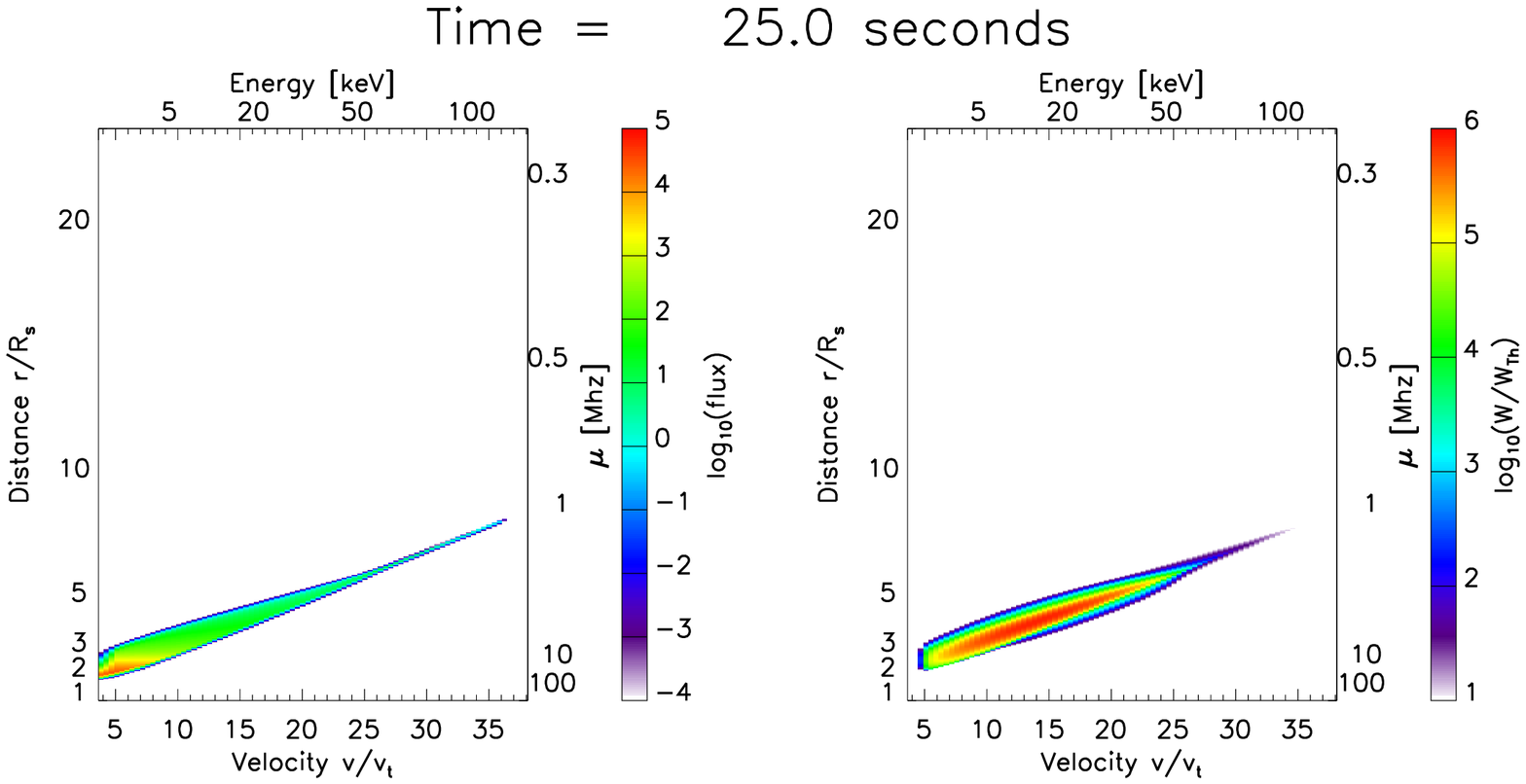}
\includegraphics[width=0.99\columnwidth]{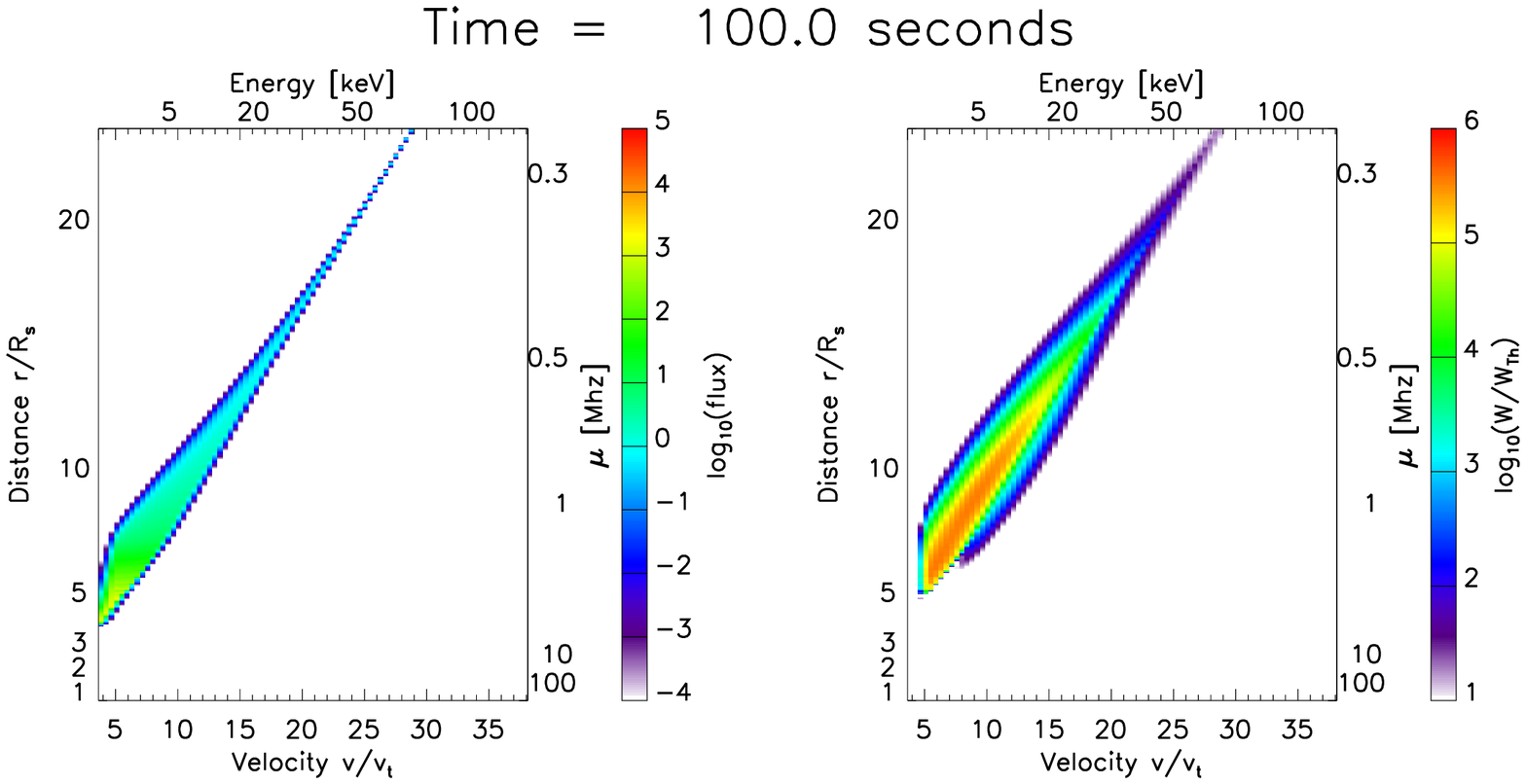}
\caption{Colour coded plot of the electron flux [cm$^2$ eV s]$^{-1}$ and spectral energy density (normalised by thermal level $W(v,x,t=0)$) of plasma waves for two moments of time.  Distance and velocity are normalised by solar radii and thermal velocity respectively.  The background plasma density is unperturbed.}
\label{fig:unpert_movie_1} \end{figure}

The initial electron distribution injected into the simulation is stable at $t=0$ but once the electrons are allowed to propagate through space, the distribution quickly becomes unstable ($\partial f/\partial v >0$) to plasma wave generation.  This is related to the `time of flight' effect first introduced by \citet{FilbertKellogg1979}.  As the growth rate of plasma waves is velocity dependent, the initial power law distribution causes quasilinear relaxation to be important up to a certain velocity or corresponding break energy. Above this energy electrons are too dilute to generate any plasma waves and travel scatter free.  Below this energy plasma waves are generated, relaxing the distribution function to a plateau in velocity space ($\partial f/\partial v \approx 0$) as energy from the electrons is transferred to the generated plasma waves. The instability forms a beam-plasma structure \citep{Melnik1995,Kontar_etal1998}, between the electron beam and the corresponding induced plasma waves seen in Figure \ref{fig:unpert_movie_1}.  We note that the presence of non-thermal particles leads to spontaneous generation of waves even when $\partial f/\partial v <0$.  However, the level of spontaneously generated waves is low and insufficient to change the electron distribution function at substantially.

The electron beam travelling through an unperturbed, decreasing background plasma experiences a slowly varying small density gradient. This decreasing gradient causes generated plasma waves to drift to smaller phase velocities as $L$ is strictly negative.  The wave energy at one point in space is thus redistributed over a wider range of phase velocities.  Therefore some generated waves are taken out of resonance with the energetic electrons which induced them.  Energetic electrons arriving later in time to this point in space are unable to reabsorb all the energy previously converted to plasma waves.  Consequently over time, the total energy in the electron beam is decreased resulting in a deceleration of electrons below the maximum energy at which plasma waves are induced.  As it was shown by \citet{KontarReid2009}, this effect leads to the formation of a broken power-law in fluence often detected at 1~AU \citep{Lin1985,Krucker_etal07}. The waves shifted towards lower velocities are eventually absorbed by the background thermal plasma via Landau damping. The recent survey of in-situ measured impulsive solar energetic electron events \citep{Krucker_etal09} suggests the break energies generally appear in the deca-keV range, in line with numerical simulations by \citet{KontarReid2009}.

\section{Electron transport through plasma with density perturbed by a sine-wave}

The background electron density is a simplified model where only large scale radial expansion is taken into account. The real inner heliosphere electron density has density fluctuations present at various smaller scales.  To initially explore density fluctuations, a simple perturbation of the background plasma is added to the previous heliospheric density model. This perturbation takes the form of a sinusoid giving a new background density
\begin{equation}\label{pert1}
n(r) = n_0(r)[1 + \alpha \sin( 2\pi r/\lambda)]
\end{equation}
where $\alpha$ and $\lambda$ are the amplitude and wavelength of the perturbation respectively and $n_0$ is the original unperturbed density.  The initial value of the amplitude $\alpha$ is taken as $10^{-2}$ while the wavelength $\lambda$ is taken as $10^{10}$~cm.  These values create a perturbation which is within reasonable solar wind parameters \citep{Celnikieretal1983}.  

\subsection{Distributions close to the Sun}

\begin{figure} \center
\includegraphics[width=0.99\columnwidth]{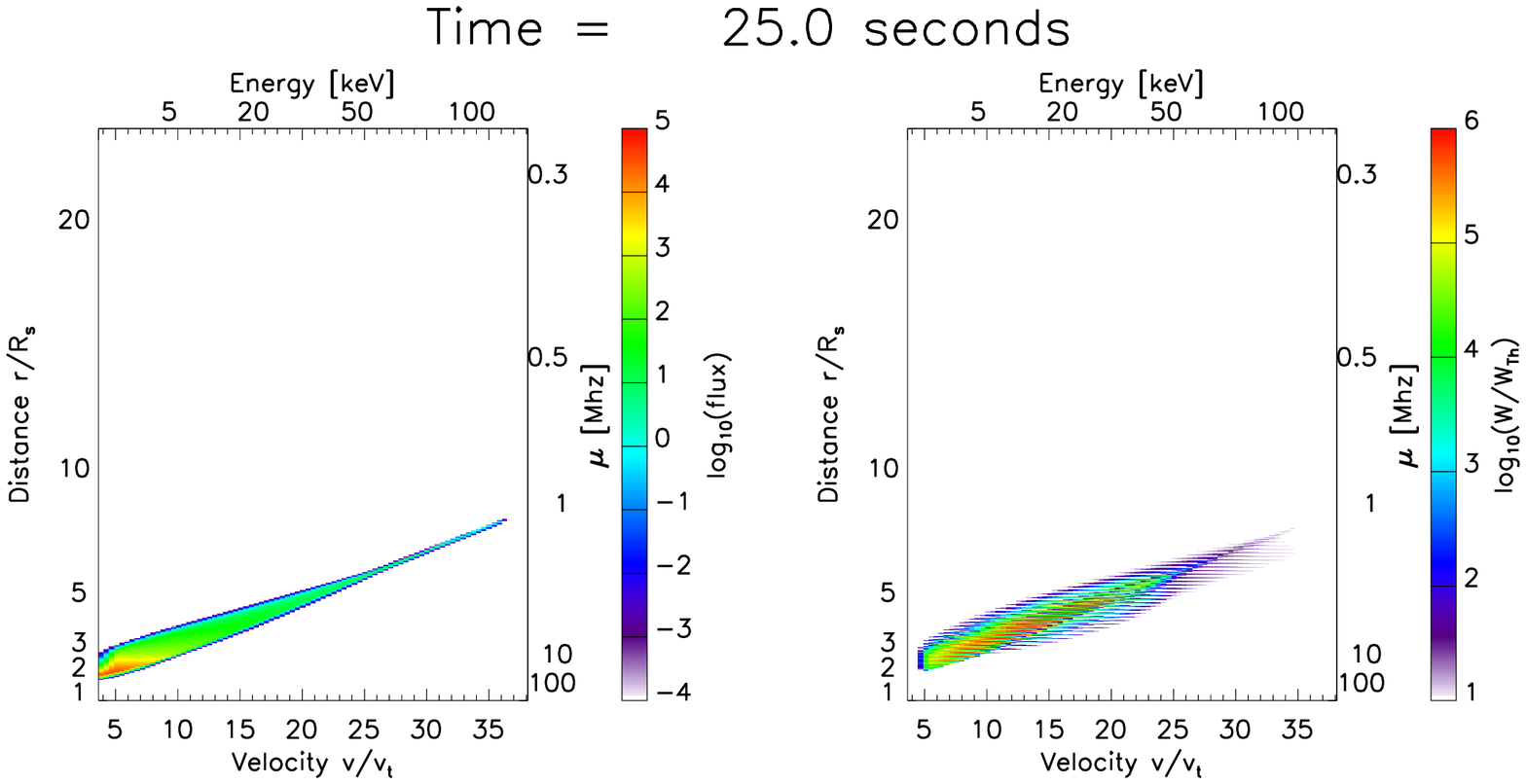}
\includegraphics[width=0.99\columnwidth]{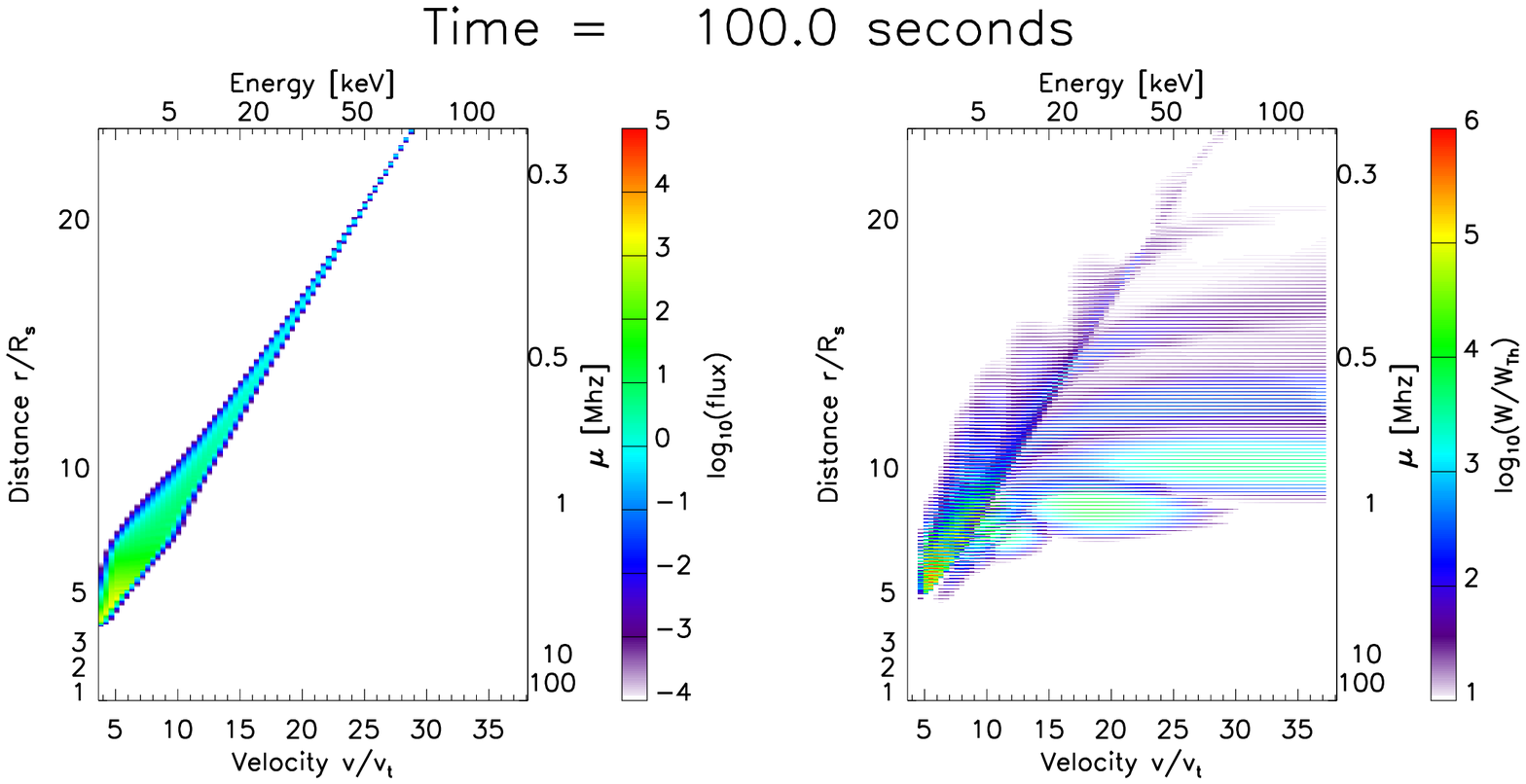}
\caption{Colour coded plot of the electron flux [cm$^2$ eV s]$^{-1}$ and spectral energy density (normalised by thermal level $W(v,x,t=0)$) of plasma waves for two moments of time.  Distance and velocity are normalised by solar radii and thermal velocity respectively.  The background plasma density has been perturbed with a sine wave.}
\label{fig:pert_movie_1} \end{figure}

Close to the Sun, the radial drop of density is very sharp and plays the dominant role in density change. The small scale fluctuations are thus unable to generate any positive density gradients. The drift of waves in velocity space is always to lower phase velocities which can be observed at the earlier time interval $t=25$~s (Figure \ref{fig:pert_movie_1}). The density fluctuations cause an increase or decrease in this drift of plasma waves to lower phase velocities. As the growth rate of plasma waves depends linearly upon the magnitude of plasma waves at any point in phase space, if the plasma inhomogeneity is too large then waves are shifted too fast and plasma wave production is suppressed \citep[in line with][]{SmithSime1979,Muschiettietal1985,Kontar2001_a,Ledenev04,Li_etal2006b}.  

\begin{figure} \center
\includegraphics[width=0.99\columnwidth]{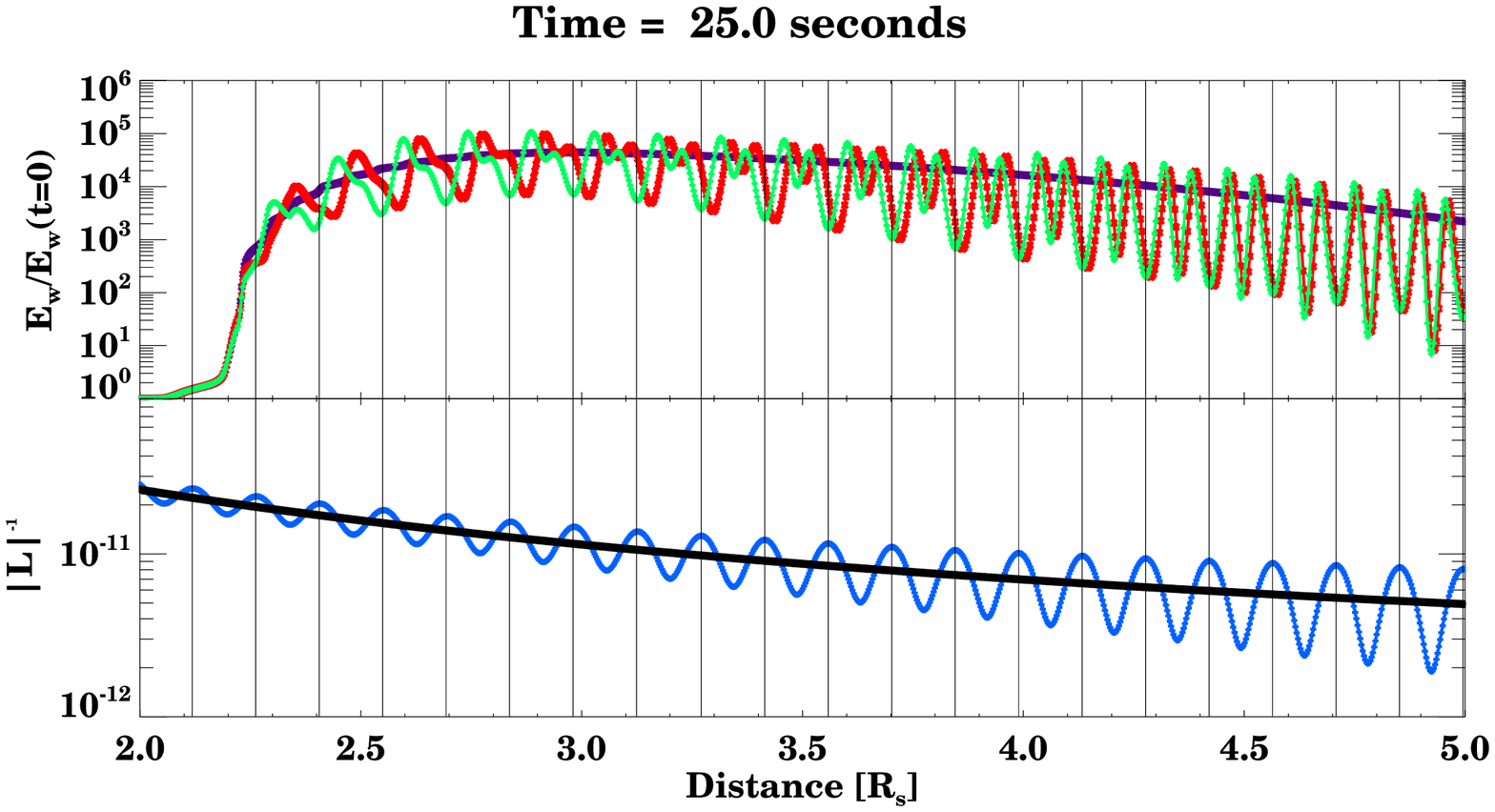}
\includegraphics[width=0.99\columnwidth]{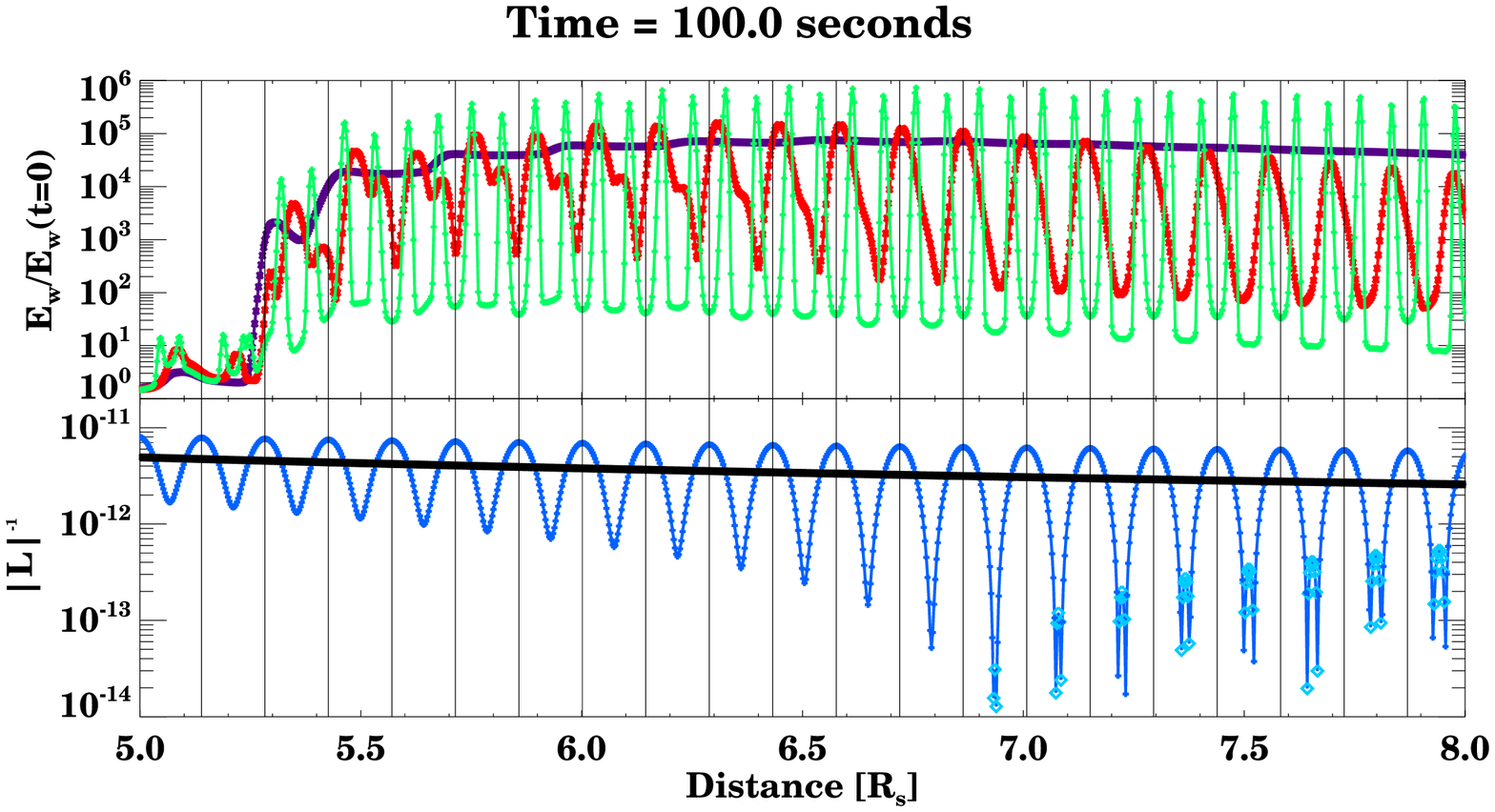}
\caption{The plasma wave energy density $E_w(r)$ at two different times for background plasma which is unperturbed (purple), perturbed (red) and perturbed without implementing group velocity (green).  The corresponding magnitude of plasma inhomogeneity $|L|^{-1}$ for unperturbed (black) and perturbed (blue) is plotted for comparison. The light blue diamonds are where the plasma inhomogeneity is positive in magnitude.}
\label{fig:pert_wave_energy_1} \end{figure}

To compare the background plasma inhomogeneity with the level of plasma waves in any spatial location we consider the magnitude of wave energy density, found by
\begin{equation}
E_w(r,t) = \int_{0}^{\infty}Wdk = \omega_{pe}\int_{v_{min}}^{v_{max}}\frac{W(r,v,t)}{v^2}dv.
\end{equation}
The plasma wave energy density, $E_w(r,t)$, close to the Sun at time $t=25$~s is displayed in Figure \ref{fig:pert_wave_energy_1} with the corresponding scale of the background plasma inhomogeneity. The unperturbed case has been over plotted for comparison. Lines have been drawn to indicate the $10^{10}$~cm wavelength of sinusoid perturbation to the background plasma.  Periodic oscillation of the background plasma is evident together with the corresponding periodic nature of the plasma wave energy density. The magnitude of $E_w(r,t)$ in the unperturbed case is generally larger than the perturbed case, showing clearly the reduction in wave growth when the background plasma is significantly perturbed. As we get further away from the Sun ($5R_s$ compared with $2R_s$) the radial drop of density plays a less dominant role allowing small scale fluctuations to become more important, seen in $|L|^{-1}$. With this increased role, the small scale fluctuations increase the suppression of induced plasma wave energy density with respect to the unperturbed case.  

Despite fluctuations suppressing plasma waves, the perturbed case displays plasma wave energy density greater than the unperturbed case at peaks in its oscillation.  The instability of the electron beam which induces the plasma waves ($\partial f/\partial v > 0$) is not fully relaxed to thermal velocities in areas of space where plasma wave production is suppressed.  Another striking feature of Figure \ref{fig:pert_wave_energy_1} is the double peak and trough behaviour of $E_w(r,t)$ within one wavelength of background plasma fluctuation.

The distribution of $E_w(r,t)$ in space is substantially different at the latter time of $t=100$~s, shown in Figure \ref{fig:pert_wave_energy_1}.  There is a larger discrepancy in magnitude between the unperturbed and perturbed case.  Moreover, the second peak of $E_w(r,t)$ within one wavelength clearly seen at $t=25$~s is suppressed at the later time of $t=100$~s.  The one remaining pronounced peak does not stay co-spatially with the small scale fluctuation wavelength but shifts backwards with respect to increasing distance from the Sun for this single point in time.  Density fluctuations at distances $\approx 7R_s$ become influential enough over the radial density decrease to generate some positive background density gradients. A positive gradient causes plasma waves to move to higher phase velocities and causes the streaking seen at $t=100$~s in Figure \ref{fig:pert_movie_1}.  Despite the plasma wave distribution being substantially different, the electron flux remains almost unchanged, agreeing with the numerical results from \cite{Kontar2001_a}. 

\subsection{The role of Langmuir wave group velocity}

The group velocity of plasma waves, $3v_{Te}^2/v$, is small in magnitude, within the range $4\times 10^7$~cm/s to  $4\times10^8$~cm/s.  At $t=25$~s (Figure \ref{fig:pert_wave_energy_1}) the removal of the group velocity term has minimal effect.  Waves are moved in space by a small distance dependent upon the magnitude of the group velocity.  The slower energetic electrons at the back of the beam produce waves with higer group velocity and hence the wave energy density is displaced further.

At the later time of $t=100$~s, $E_w(r,t)$ is substantially different when the group velocity term is not present, seen in Figure \ref{fig:pert_wave_energy_1}.  There is a clear double peak and trough behaviour within one background density fluctuation wavelength.  Without any group velocity, waves are unable to travel from points in space where the background density structure favours wave growth to points where wave growth is suppressed.  The simulation with no group velocity also has a higher magnitude of wave energy density at its peaks than both the other simulations.

The group velocity of plasma waves, despite being small, acts to move wave energy from points in space where plasma waves are strongly induced to points in space where they are suppressed.  This has a spatial smoothing effect on the induced plasma wave energy density.

\subsection{Amplitude of fluctuations}

The amplitude $\alpha$ of the density fluctuations directly varies the background electron plasma density. The magnitude of this factor near the Earth can be found from observational results. It has been measured using the International Sun-Earth Explorer (ISEE) \citep{Harvey_etal1978} propagation experiment with scintillation techniques \citep{Celnikieretal1983} that the background electron plasma density near the Earth varies by about $10\%$. This would give the amplitude of $\alpha \le 0.1$.  Therefore we consider $\alpha$ in the range $10^{-3} \le \alpha \le 10^{-1}$. The wavelength of the perturbation was taken as $\lambda=10^{10}$~cm.

As $\alpha$ increases in magnitude, the oscillation in wave energy density increases. Similarly as $\alpha$ decreases in magnitude, the oscillations in wave energy density decrease such that as $\alpha \rightarrow 0$ the wave energy density tends to the state where no perturbations are present in the background electron plasma density.  This can be seen in Figure \ref{fig:WE_A_L_1} in the plasma inhomogeneity, $|L|^{-1}$. As $\alpha$ decreases to $10^{-3}$, the plasma inhomogeneity does not vary as much and $L$ remains negative.

The variation of $\alpha$ in Figure \ref{fig:WE_A_L_1} shows how the magnitude of the plasma inhomogeneity affects wave generation. If the fluctuations are too large, plasma waves drift in phase velocity too fast from the beam and are unable to build up. This suppression can clearly be seen when $\alpha=10^{-1}$. Most spatial areas have large values of $|L|^{-1}$ and corresponding low values of wave energy density.  Conversely, when $\alpha=10^{-3}$, the small scale fluctuations are small and wave energy density is able to build up to high magnitudes. This suppression agrees with previous theoretical \citep{Melrose1982,Melroseetal1986} and numerical work on Langmuir wave generation in non-uninform plasma \citep{Kontar2001_a}.

\subsection{Wavelength of perturbations}

The wavelength of density fluctuations $\lambda$ has a strong effect on the local scale of plasma inhomogeneity, $L$, through $dn/dr$ having one term inversely proportional to $\lambda$.  Density fluctuations have been measured at a variety of different length scales from $10^{12}$~cm down to $10^6$~cm \citep{Neugebaueretal1978,Celnikieretal1987,Kelloggetal2009}.  We have varied $\lambda$ in the range $10^{9}$~cm $\le \lambda \le 10^{11}$~cm which is close to the range of fluctuations presented by \citet{Celnikieretal1987}.  The amplitude was set to $\alpha=10^{-2}$, similar to the previous section for comparison reasons.

\begin{figure} \center
\includegraphics[width=0.79\columnwidth]{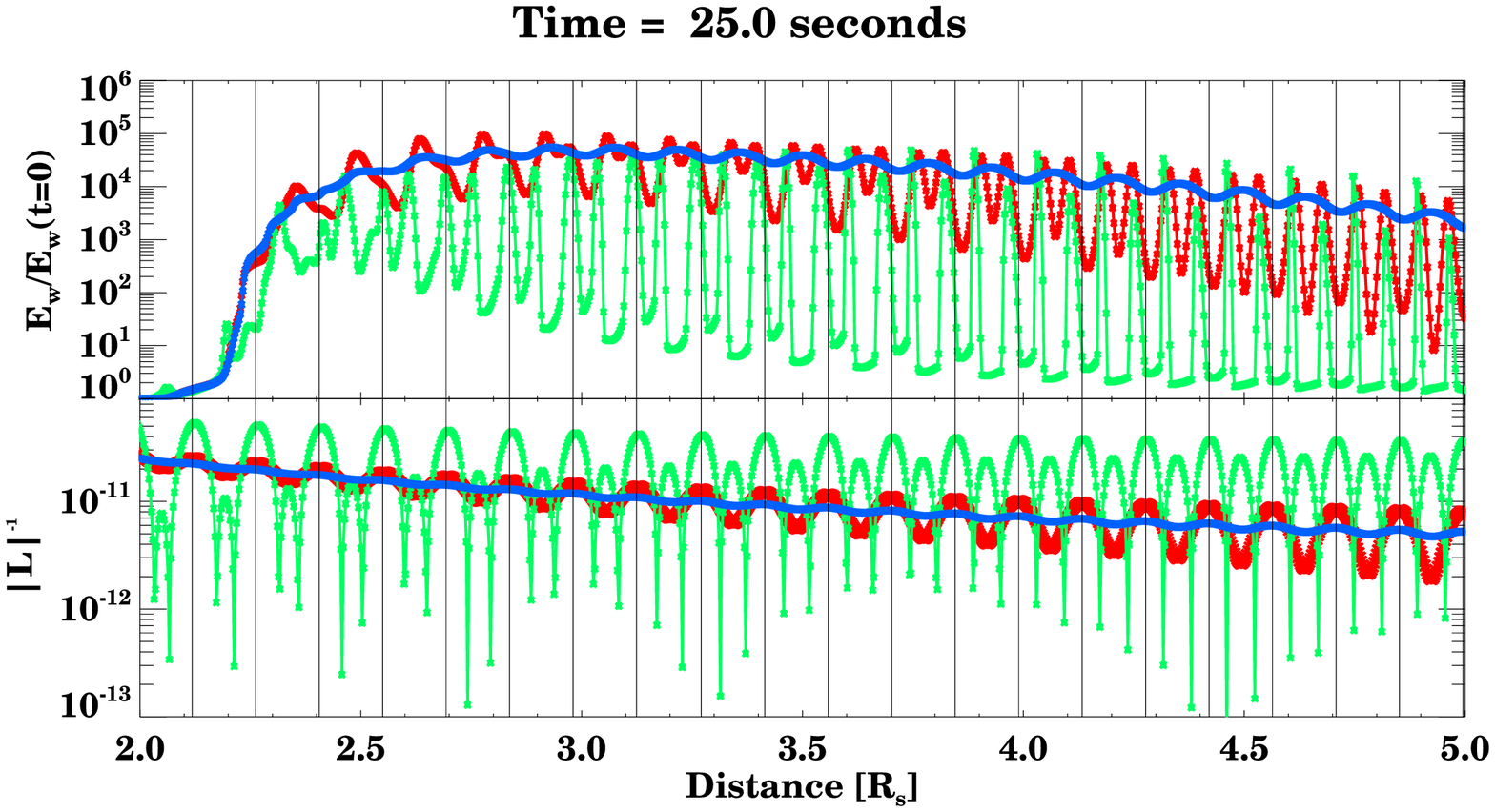}
\includegraphics[width=0.79\columnwidth]{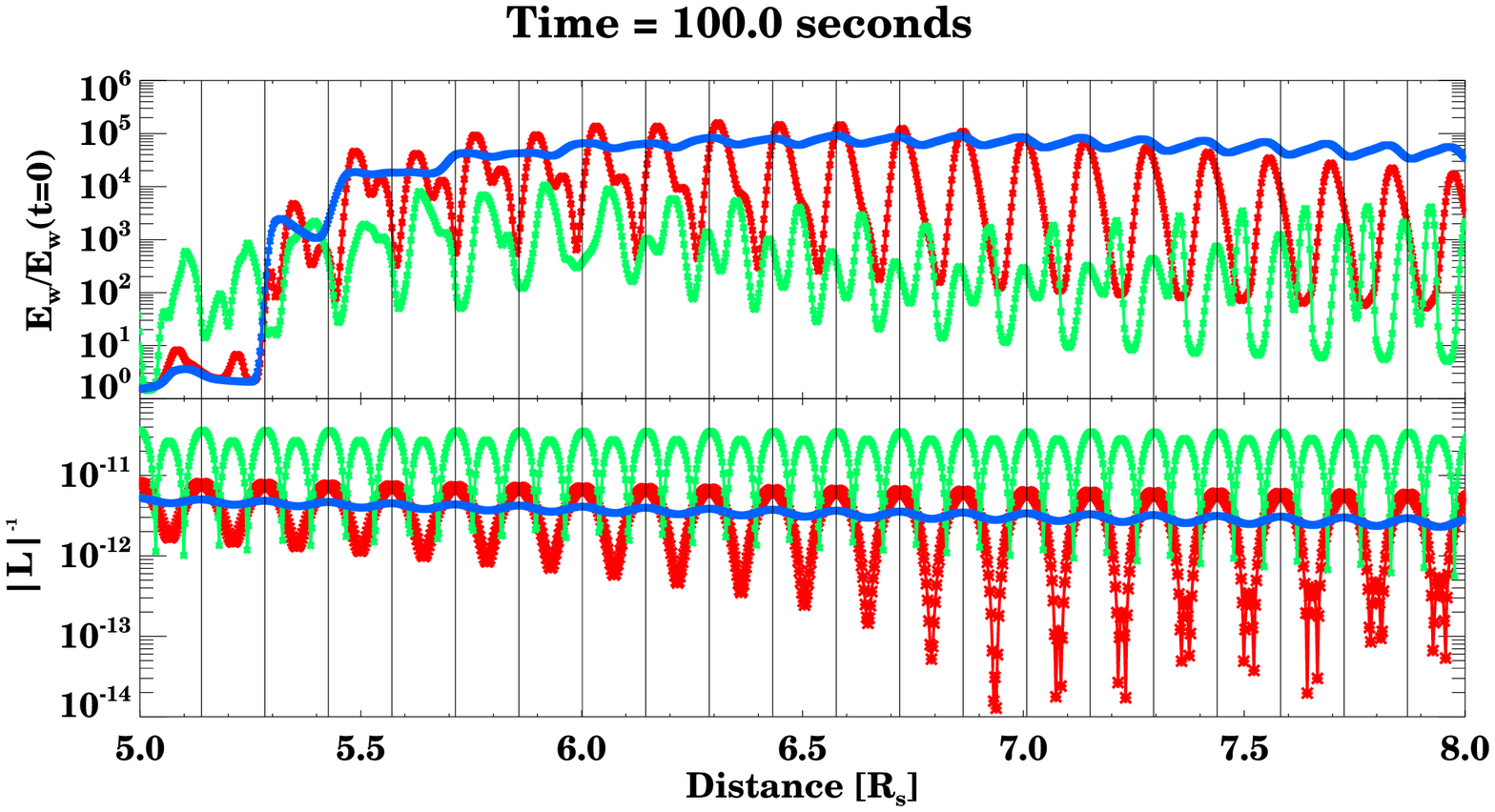}
\includegraphics[width=0.79\columnwidth]{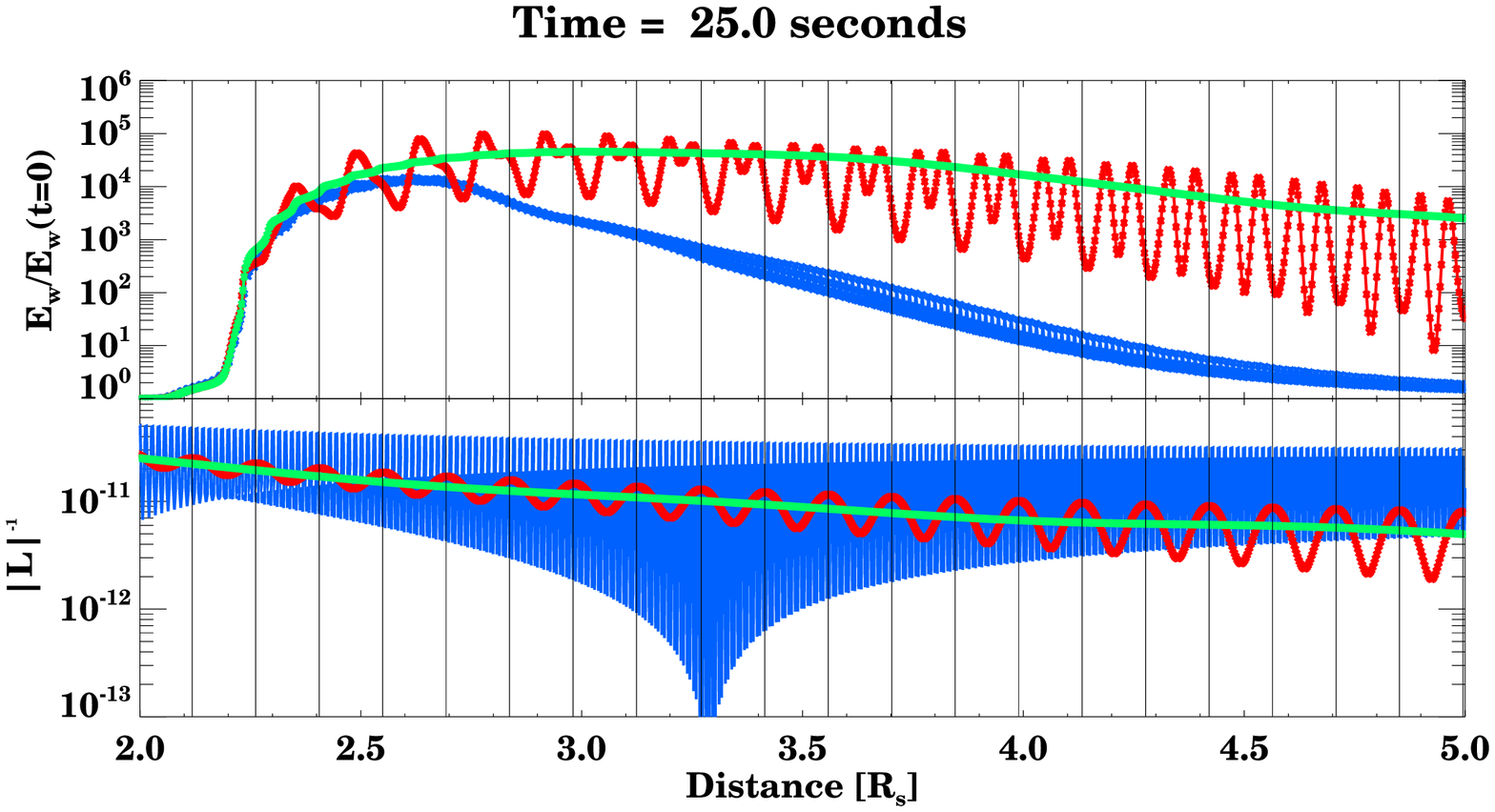}
\includegraphics[width=0.79\columnwidth]{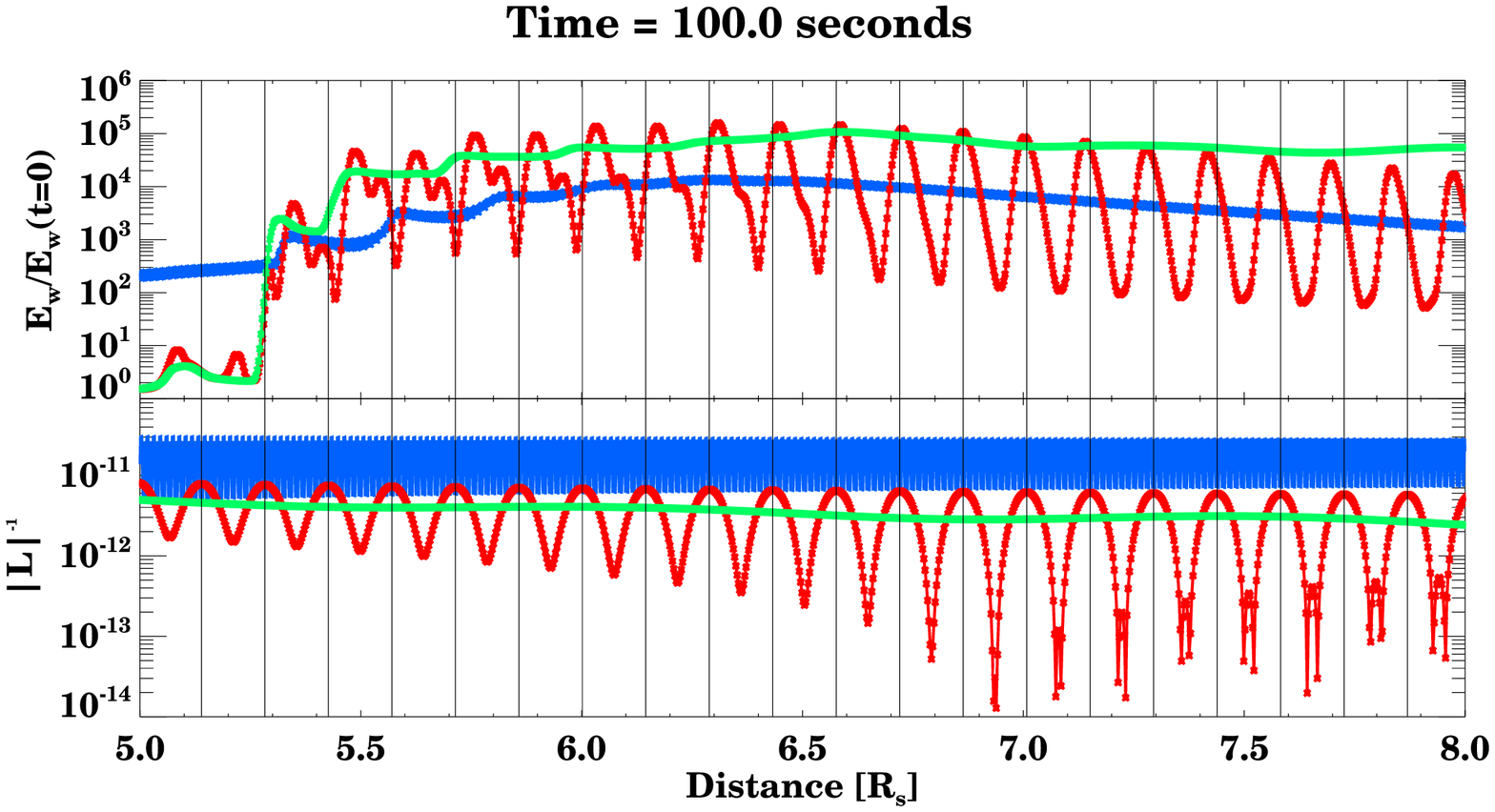}
\caption{Top: Energy density of Langmuir waves for $\alpha = 10^{-1}$ (green), $10^{-2}$ (red), $10^{-3}$ (blue). $\lambda=10^{10}$~cm.  Bottom: Wave energy denisty for $\lambda = 10^{11}$~cm(green), $10^{10}$~cm(red), $10^{9}$~cm(blue).  $\alpha=10^{-2}$.  The background plasma inhomogeneity $L(r)$ for each simulation in the appropriate colour is shown in lower panels.}
\label{fig:WE_A_L_1} \end{figure}

As $\lambda$ increases in magnitude, the oscillation in density inhomogeneity decreases such that as $\lambda \rightarrow \infty$, the wave energy density tends to the state where no perturbations are present in the background electron plasma density.  This can be seen from Figure \ref{fig:WE_A_L_1} in the case where $\lambda=10^{11}$~cm and the density inhomogeneity is very smooth.  Conversely, as $\lambda$ decreases, the magnitude of $L^{-1}$ increases while the sign of $L^{-1}$ fluctuates rapidly.

We can see from Figure \ref{fig:WE_A_L_1} that when $\lambda$ is large, the induced plasma wave energy density resembles the unperturbed case.  When $\lambda$ is small, the large magnitude of $L^{-1}$ causes waves to shift in velocity space faster.  At any spatial point waves are present with a far greater range of phase velocities, however, their magnitude is much decreased.  This means there exists a decreased level of plasma waves at points in phase space where the electron beam is present.  The growth factor of plasma waves, responsible in the kinetic equations for converting electron beam energy to plasma wave energy, is proportional to the level of plasma waves.  The decreased level of plasma waves in areas of phase space where the electron beam is present causes less energy to be transferred from electron beam to plasma wave and is the reason for the smaller wave energy density observed in Figure \ref{fig:WE_A_L_1} when $\lambda=10^9$~cm.

\section{Electron transport through plasma with power-law density fluctuations}

The power spectrum of density fluctuations observed in the solar wind density follows a simple, Kolmogorov type power law near the Earth with spectral index near to $5/3$.  A similar spectrum index of perturbations has been observed both with scintillation techniques \citep{Celnikieretal1983,Celnikieretal1987} and with in-situ measurements \citep{Neugebaueretal1978,KelloggHorbury2005}.  The spectrum has been observed to steepen at small wavenumbers around $10^{8}$~cm.  To model small scale density fluctuations many perturbations of the background plasma are introduced, so the density is
\begin{equation}\label{pertm1}
n(r) = n_0(r)\left[1 + C \sum_{n=1}^{N}  \lambda_n^{\beta/2} sin( 2\pi r/\lambda_n + \phi_n)\right]
\end{equation}
for N perturbations where $n_0(r)$ is the original unperturbed density. $\lambda_n$ are the wavelengths of density perturbations with $\phi_n$ as their random phase. $C$ is a constant which normalises the density fluctuations given by
\begin{equation}\label{pertm2}
C = \sqrt{\frac{2\langle \Delta n(r)^2\rangle}{\langle n(r)\rangle^2 \sum_{n=1}^{N}\lambda_n^{\beta}}}
\end{equation}
where $\langle n(r)\rangle$ is the mean density.  The root mean squared deviation of the density , $\sqrt{\langle \Delta n(r)^2\rangle}$, from observational values near the Earth was taken to be $0.4$~cm$^{-3}$ or $10\%$ of the mean density.  The quantity $\sqrt{\frac{\langle \Delta n(r)^2 \rangle}{\langle n(r) \rangle^2}}$, the fractional density fluctuations, is a measure of the turbulent intensity of the background plasma.  From Equation \ref{pertm2} this quantity is radially independent giving a constant turbulent intensity from the Sun to the Earth.  We can then model the radial variation of turbulent intensity with 
\begin{equation}\label{pertm3}
\sqrt{\frac{\langle \Delta n(r)^2 \rangle}{\langle n(r) \rangle^2}}
 = \left(\frac{n_0(1AU)}{n_0(r)} \right)^{\psi} \sqrt{\frac{\langle\Delta n(r_{1AU})^2 \rangle}{\langle n(r_{1AU}) \rangle^2}}
\end{equation}
where $\psi \ge 0$ determines the degree at which the density fluctuations become less dominant with $\psi=0$ correspoding to no radial variation.  For simplicity, we will reference the fractional density fluctuations as $\Delta n/n$.  We considered the range on $\lambda$ to be $10^{7}$~cm $\le \lambda \le 10^{10}$~cm which is within the inertial range of solar wind turbulence.  Larger values of $\lambda$ have a minor effect and the amplitude of waves shorter than $\lambda\approx10^7$~cm is small.  The random phases $0 \le \phi < 2\pi$ ensure the amplitudes of density fluctuations have a Gaussian distribution.

\begin{figure} \center
\includegraphics[width=0.79\columnwidth]{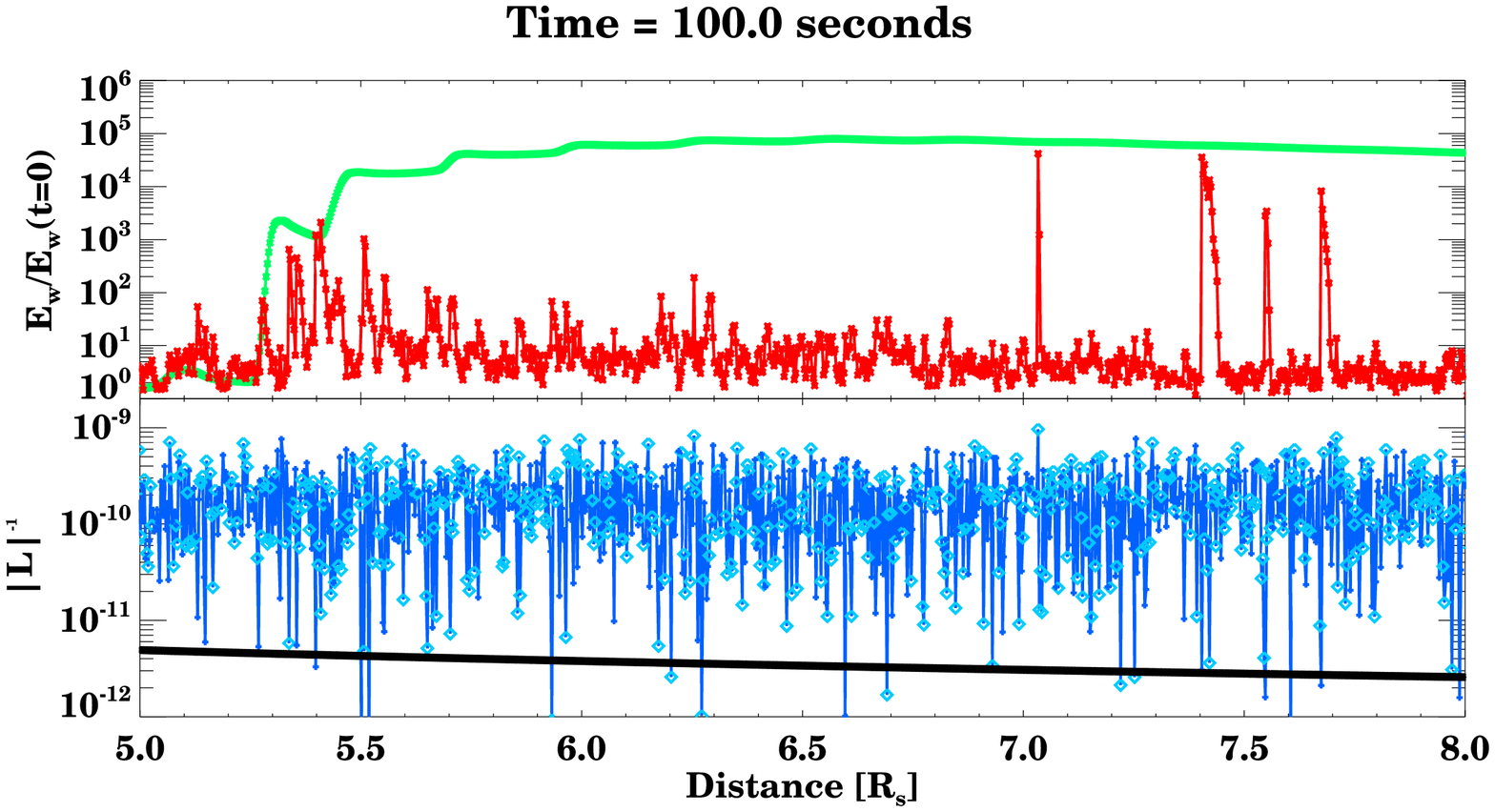}
\includegraphics[width=0.79\columnwidth]{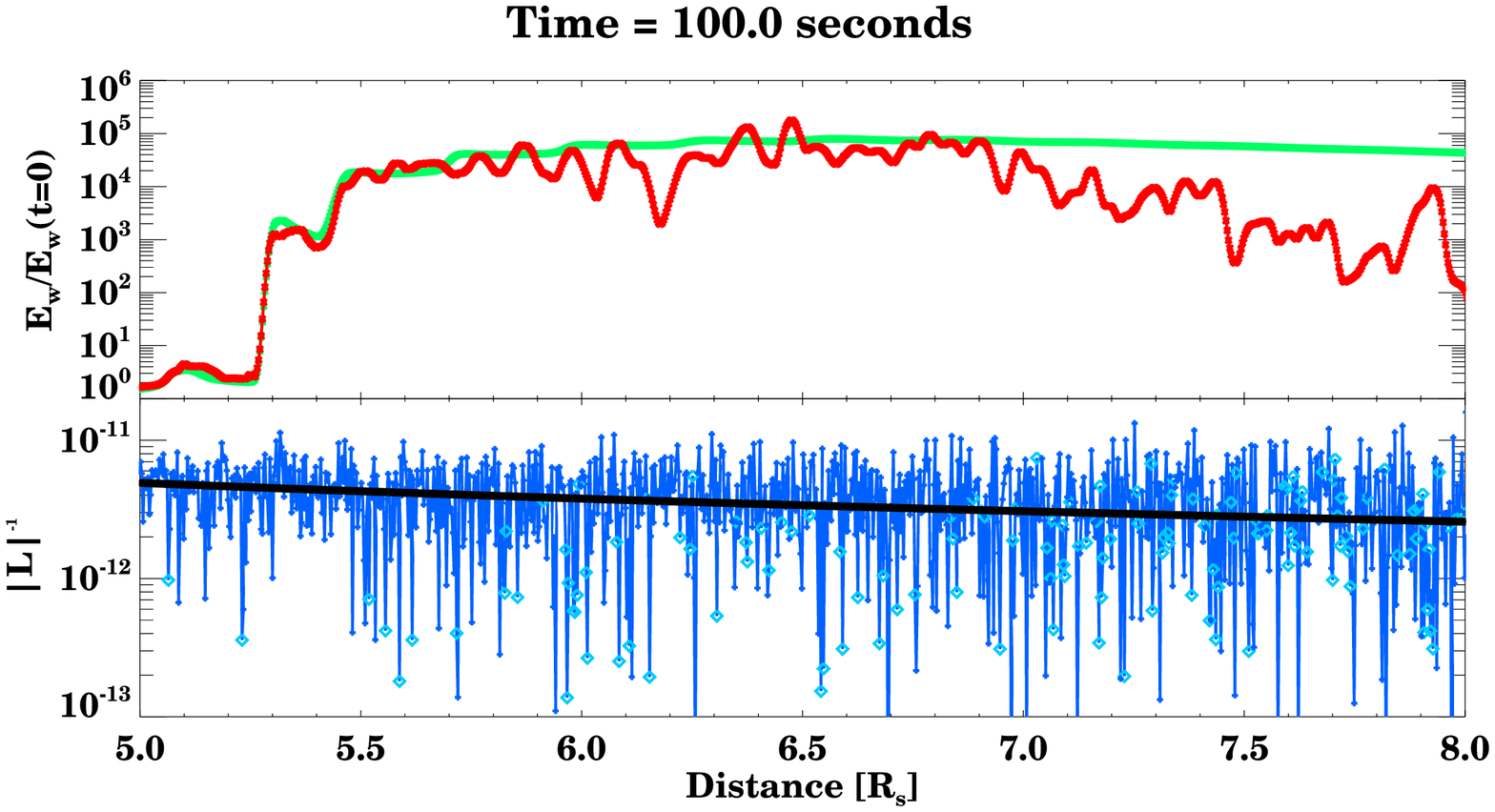}
\caption{The plasma wave energy and corresponding plasma inhomogeneity when density fluctuations have a power law spectra in frequency space and $\Delta n/n=10\%$ at the Earth  Top: The fluctuations are constant from the Sun to the Earth ($\psi=0$).  Bottom: The fluctuations increase from the Sun to the Earth ($\psi=0.5$).  Both graphs are over plotted with the unperturbed case (green).  The plasma inhomogneity is plotted for unperturbed case (black) and perturbed case (blue) with light blue diamonds for positive values.  }
\label{fig:WE_M_1} \end{figure}

A constant level of $\Delta n/n$ thoughout the inner heliosphere is found by setting $\psi=0$.  We set $\Delta n/n=10\%$ which is within the observed range of values near the Earth \citep{Celnikieretal1987}.  Figure \ref{fig:WE_M_1} shows the density inhomogeneity and corresponding plasma wave energy density close to the Sun.  The high level of inhomogeneity caused by the small scale fluctuations greatly suppresses plasma wave spatial build-up compared to the unperturbed case.  This suppression of plasma waves for $\Delta n/n=10\%$ close to the Sun can prevent the high level of plasma waves required for type III solar radio emission.

\begin{figure} \center
\includegraphics[width=0.99\columnwidth]{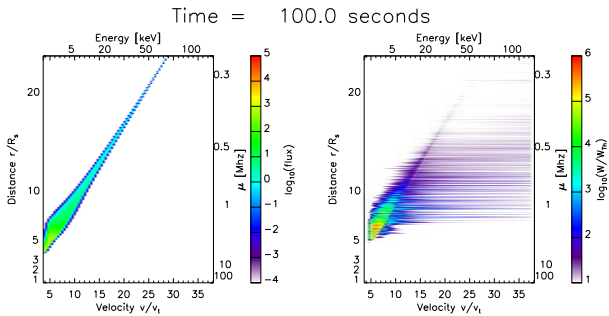}
\caption{Colour coded plot of the electron flux [cm$^2$ eV s]$^{-1}$ and spectral energy density (normalised by thermal level $W(v,x,t=0)$) of plasma waves.  Distance and velocity are normalised by solar radii and thermal velocity respectively.   The small scale fluctuations have a power law spectra in frequency space where the fluctuations increase from the Sun to the Earth with $\psi=0.5$.}
\label{fig:movie_100s_multi} \end{figure}

To vary the level of fluctuations from the Sun to the Earth, we set $\psi > 0$.  Figure \ref{fig:WE_M_1} shows the density inhomogeneity and corresponding plasma wave energy density close to the Sun with $\Delta n/n=10\%$ at 1~AU and $\psi=0.5$.  Comparing the plasma wave energy density with the earlier simulations which assumed constant $\Delta n/n$, we can see a much larger magnitude of plasma wave energy density being induced from the electron beam.  The reduced contribution from the small scale fluctuations allows build up of plasma wave energy density.  This high level of wave energy density is required to see the recorded brightness temperatures associated with type III radio emission.  Moreover, the spatial spread of plasma waves is much less sporadic than the produced level of wave energy density in the simulation with high level of fluctuations.  The electron beam and plasma wave distribution can be seen in Figure \ref{fig:movie_100s_multi}.  Plasma waves are no longer perturbed in a periodic fashion but are pseudo-random in space.  The pseudo-random nature of the spikes in plasma wave energy density leads to similar clumpy behaviour of plasma waves observed in-situ by spacecrafts \citep[e.g.][]{GurnettAnderson1976}.

\section{Electron spectra near the Earth}

Previous work \citep{KontarReid2009} has shown the generation and absorption of plasma waves coupled with the effect of the background plasma inhomogeneity can change the electron beam energy distribution. A broken power-law in fluence spectrum can be formed from an initially single power-law distribution.  The break at which the two power-laws connect is the maximum velocity the electrons were able to induce plasma waves through a resonant interaction with the background plasma.  The spectrum below the break is flattened during transport because the electron beam is unable to re-absorb all the energy transferred to plasma waves due to background plasma density gradients.

\begin{figure} \center
\includegraphics[width=0.49\columnwidth]{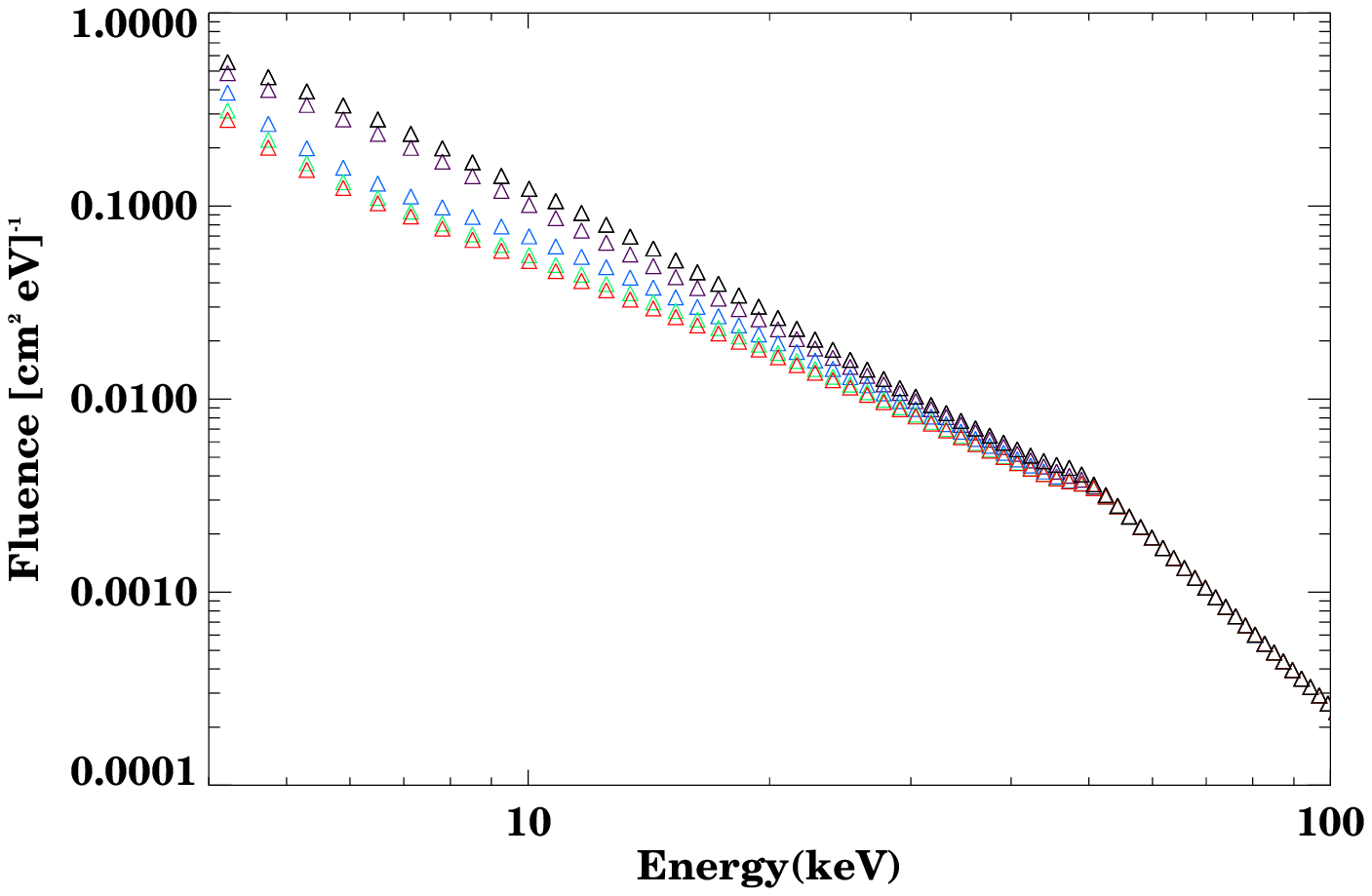}
\includegraphics[width=0.49\columnwidth]{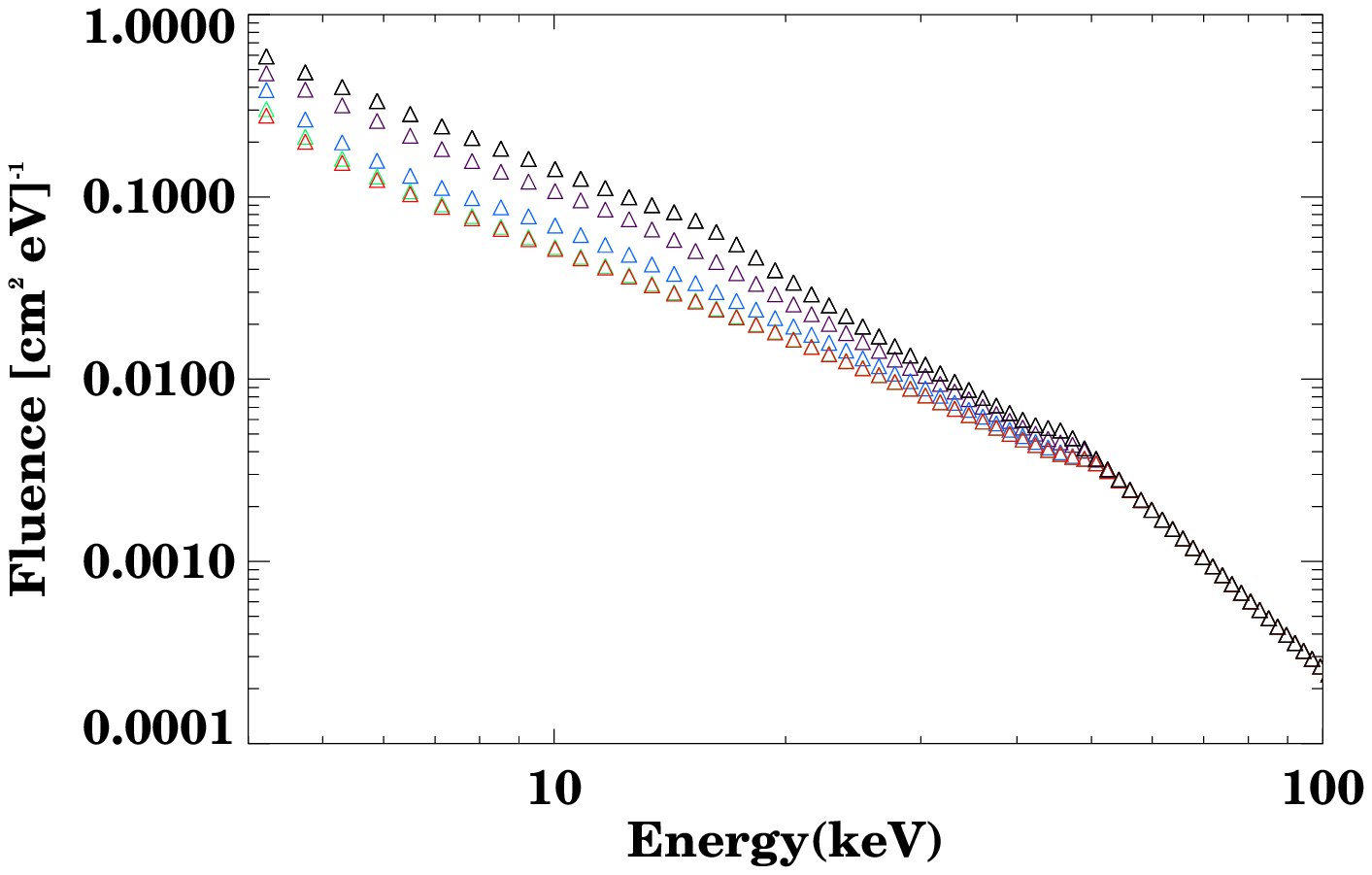}
\includegraphics[width=0.49\columnwidth]{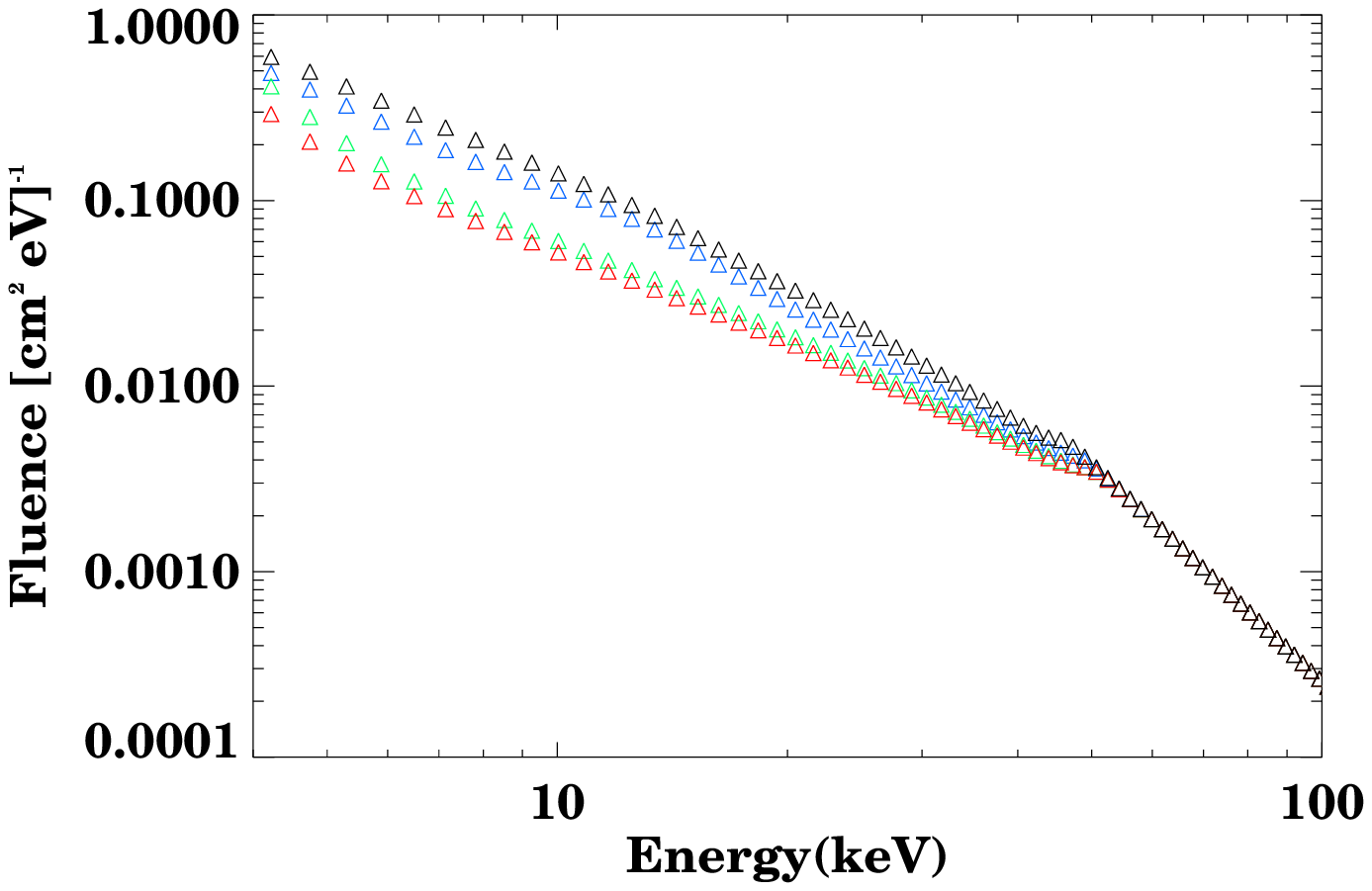}
\includegraphics[width=0.49\columnwidth]{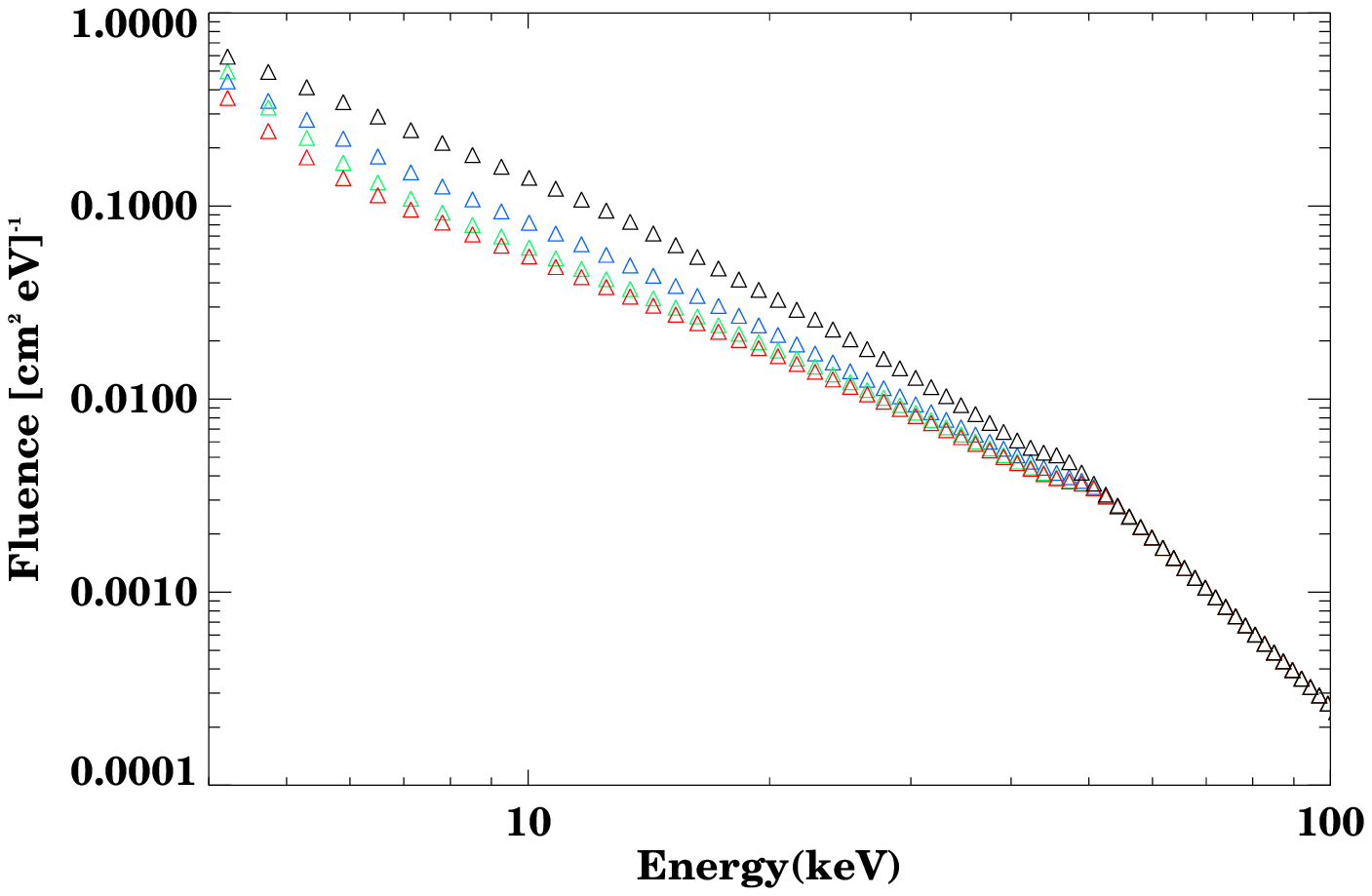}
\caption{Fluence of the electron distribution function near the Earth.  Top Left: Five simulations with $\alpha=10^{-1}$ (black), $10^{-1.5}$ (purple), $10^{-2}$ (blue), $10^{-2.5}$ (green), and $10^{-3}$ (red).  $\lambda=10^{10}$~cm.  Top Right: Five simulations with $\lambda=10^{8}$~cm (black), $10^{9}$~cm (purple), $10^{10}$~cm (blue), $10^{11}$~cm (green) and unperturbed (red). $\alpha=10^{-2}$.  Bottom Left: Four simulations for multi-scale fluctuations with $\Delta n/n$ of $10~\%$~(black), $1~\%$~(blue), $0.1~\%$~(green) and $0.01~\%$~(red) of the mean background density.  Bottom Right:  Four simulations for multi-scale fluctuations which decrease in power close to the Sun for $\psi$ of $0$ (black), $0.3$ (blue), $0.5$ (green), $0.8$ (red).}
\label{fig:PF_F_all} \end{figure}

\begin{figure} \center
\includegraphics[width=0.40\columnwidth]{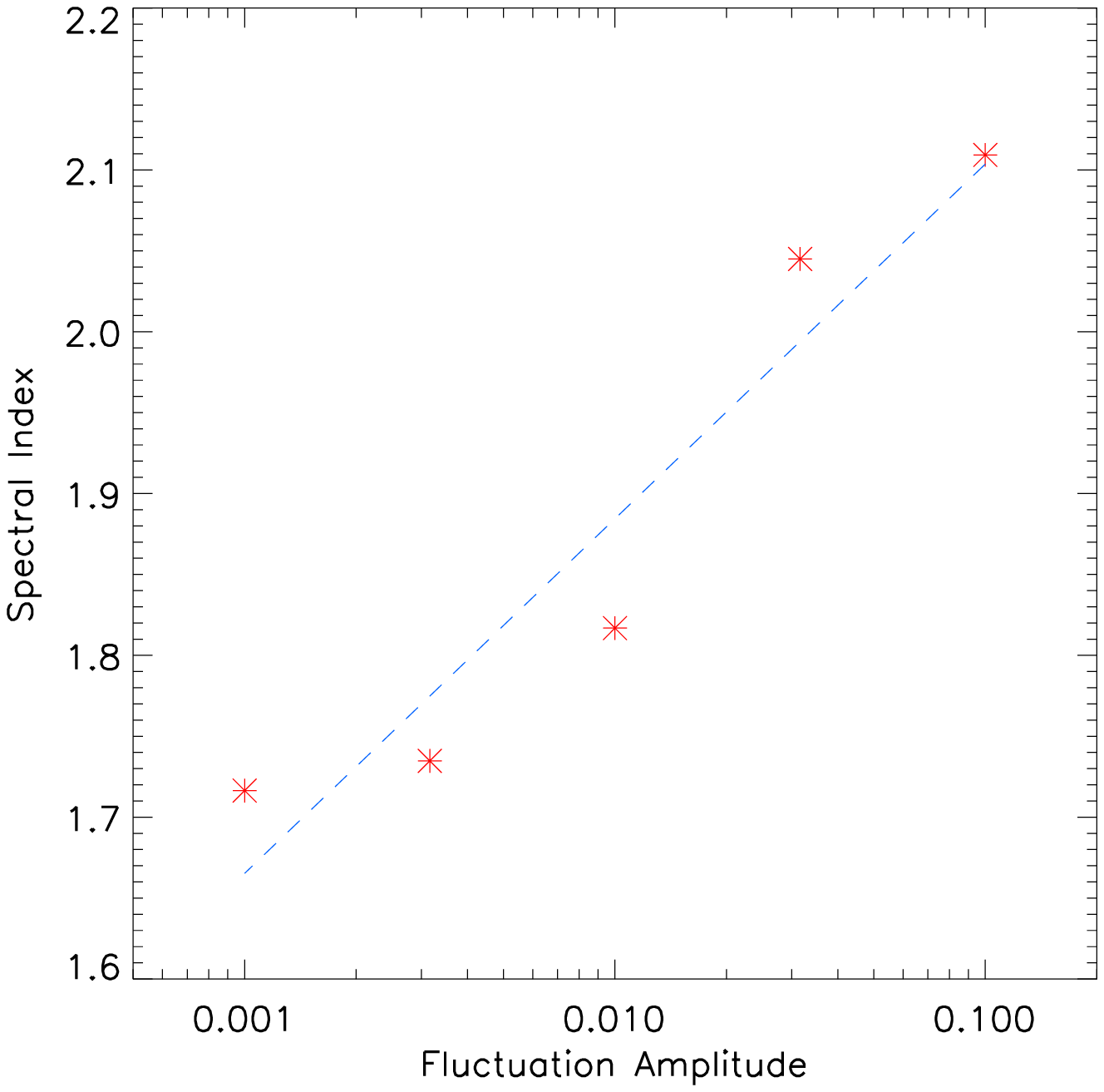}
\includegraphics[width=0.40\columnwidth]{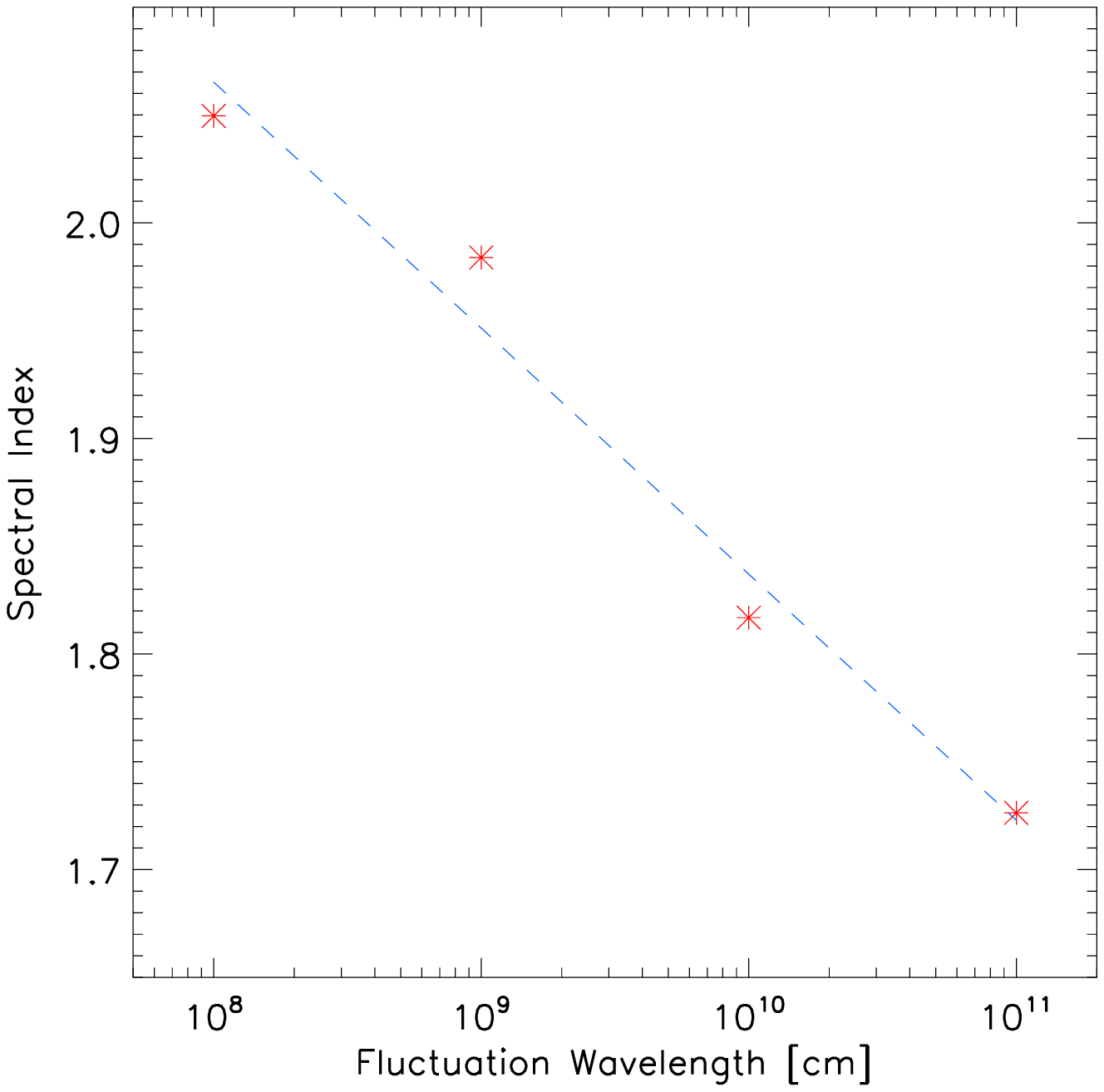}
\includegraphics[width=0.40\columnwidth]{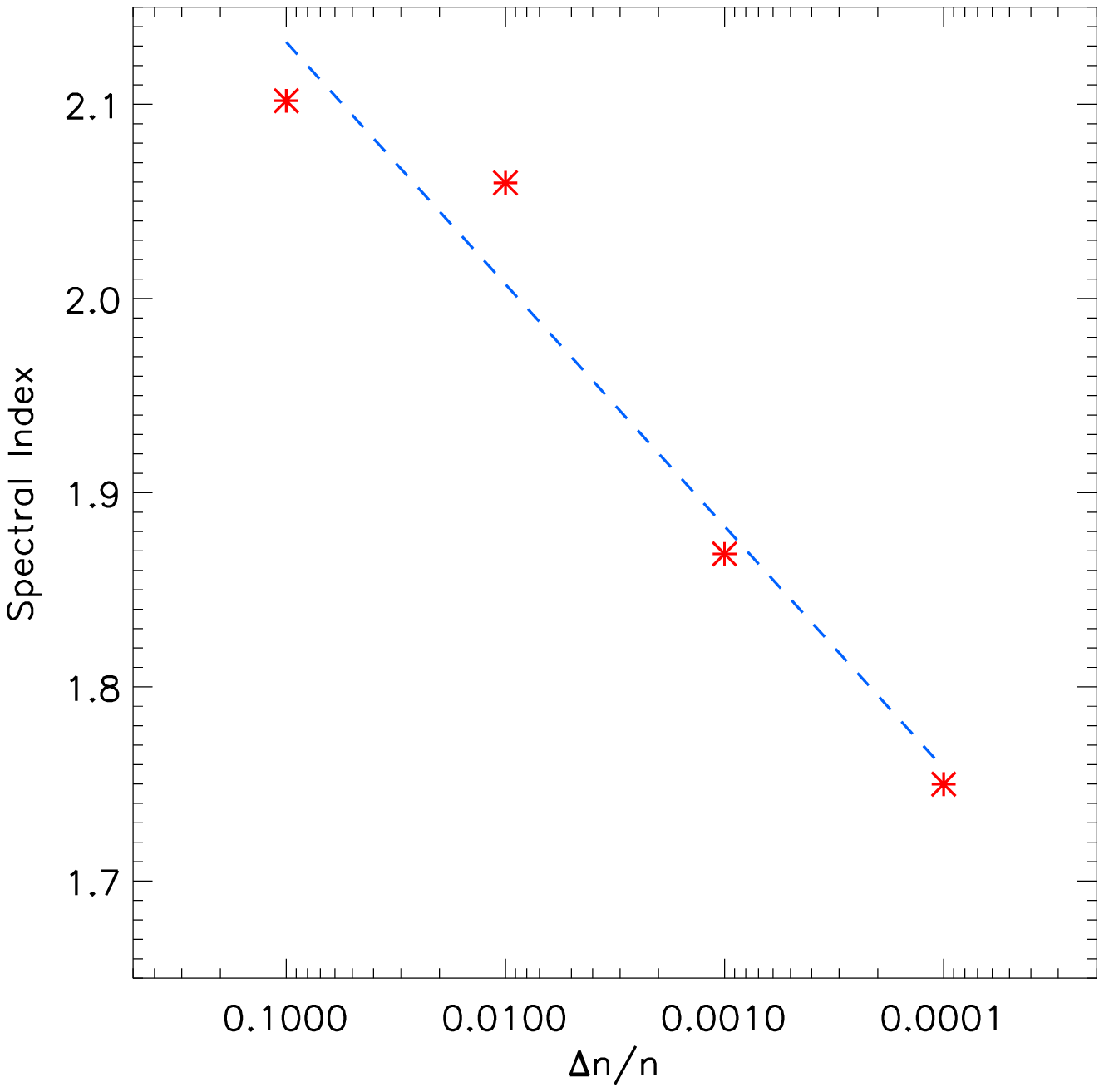}
\includegraphics[width=0.40\columnwidth]{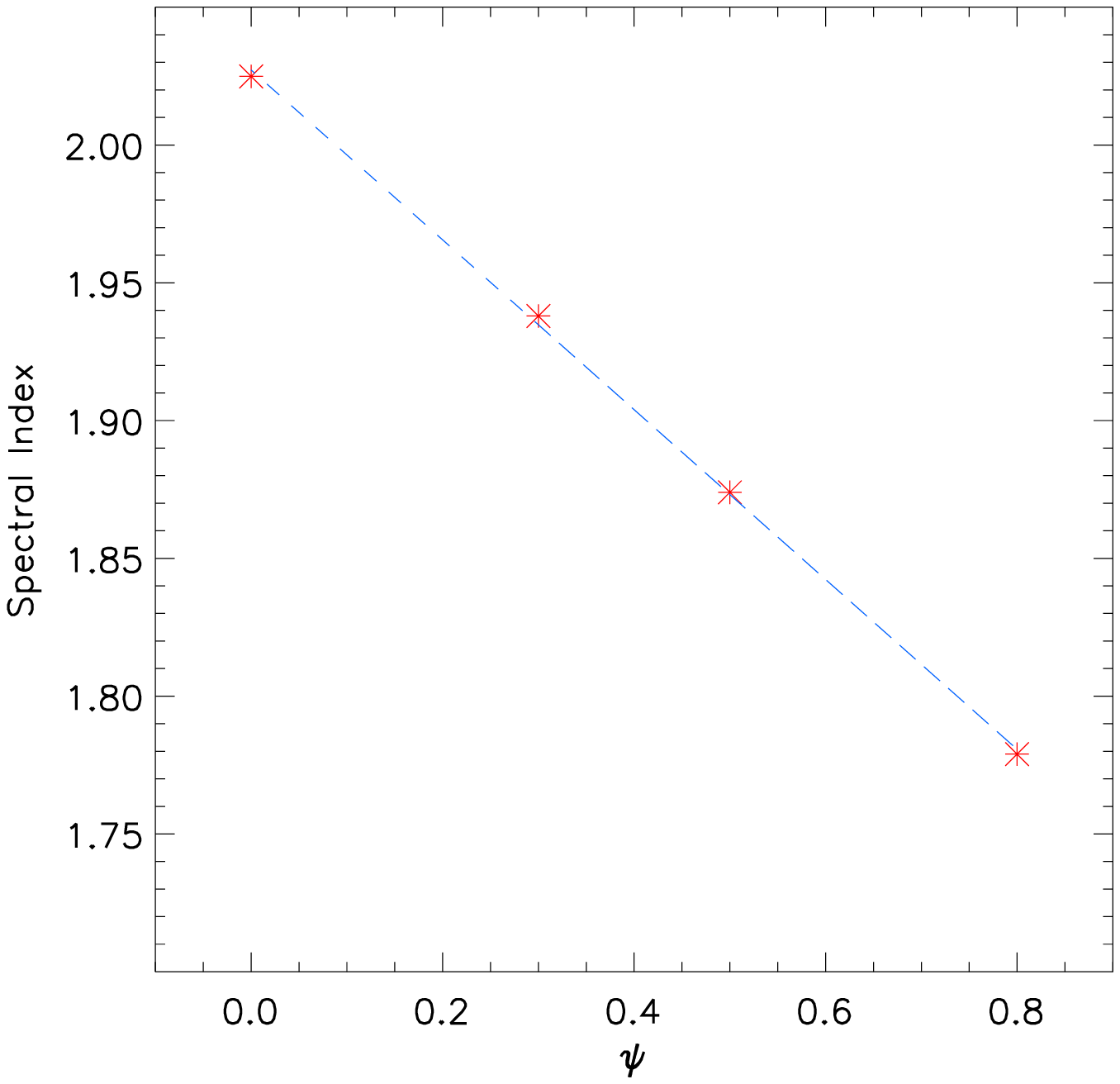}
\caption{The spectral index of a power law fit between 4 and 40 keV for the fluence spectra of electrons near the Earth. Top Left: Spectral index versus the amplitude of density fluctuation. Top Right: Spectral index versus the wavelengths of density fluctuation.  Bottom Left: Spectral index versus multi-scale level of fluctuations.  Bottom Right: Spectral index versus $\psi$, the radial degree at which density fluctuations become less dominant.}
\label{fig:SI_str_len_pert} \end{figure}

Introducing small scale density fluctuations into the background plasma changes its properties and should consequently change the spectrum of the electron distribution function.  Whilst changes in the electron spectrum are not visible on short scales (a few relaxation times), the fluence spectrum at the Earth shows a noticeable dependence upon the level of fluctuations introduced to the simulation.  Figure \ref{fig:PF_F_all} shows the fluence spectrum of the electron beam at the Earth for five different amplitudes of fluctuation within the range $10^{-3} \leq \alpha \leq 10^{-1}$.  As shown earlier, the small scale density fluctuations suppress the generation of plasma waves.  This decreases the amount of energy transferred through resonant interaction from the electron beam to the plasma waves.  With less total energy, a smaller amount of energy in plasma wave form can drift to higher or lower phase velocities and not be re-absorbed by the electron beam.  The amount of deceleration the electron beam can undergo due to plasma waves drifting is decreased, causing a reduction in the flattening of the fluence electron spectrum.  This means when $\alpha$ is larger, the fluence spectrum below the break energy has a higher spectral index (Figure \ref{fig:SI_str_len_pert}).  Similar behaviour is demonstrated by the fluence spectrum of the electron beam at the Earth for four different wavelengths of small scale fluctuations within the range $10^8$~cm $ \le \lambda \le 10^{11}$~cm, shown in Figure \ref{fig:PF_F_all}. The same lack of wave generation for small $\lambda$ reduces the deceleration the electron beam undergoes and hence reduces the flattening of the fluence spectrum (Figure \ref{fig:SI_str_len_pert}).

Despite the change in spectrum near the Earth, the electron distribution function does not share the same sensitivity to the structure of the background electron density as the plasma waves (See Figures \ref{fig:unpert_movie_1} and \ref{fig:pert_movie_1}).  The simulation with perturbed plasma does however show small changes, most noticeably in the tail of the electron distribution.  A positive spatial gradient in background plasma causes plasma waves to drift to higher phase velocities.  This drifting of waves in velocity space allows their energy to be re-absorbed by the tail of the beam such that electrons are accelerated to higher energies.  It is the opposite effect of the negative density gradient taking plasma wave energy away from electrons and forming a broken power-law near the Earth.  This acceleration of electrons causes the noticeable bump around $10-20$~keV in Figure \ref{fig:PF_F_all}, seen for simulations with higher spectral indices below the break energy.  The bump becomes more prominent for small $\lambda$, high $\alpha$ or more generally when the background density fluctuations are more effective at moving wave energy to higher phase velocities.

Extending the density  fluctuations to multi-wavelength model, a Kolmogorov type power-law is assumed where $(\Delta n/n)^2 \sim \lambda^{5/3}$ with $\Delta n/n$ remaining radially constant ($\psi=0$).  Figure \ref{fig:PF_F_all} shows the fluence spectrum at the Earth for four different turbulent intensities $0.01\% \le \Delta n/n \le 10\%$.  The larger $\Delta n/n$ is, the greater the suppression of plasma waves and hence the higher the spectral index below the break energy of the fluence spectra (Figure \ref{fig:SI_str_len_pert}).  The signature bump can be seen in the spectra around $10-20$~keV, again caused by the acceleration of electrons at the back of the beam due to plasma waves drifting to higher phase velocities.  

The electron beam fluence spectra for density fluctuations $\Delta n/n$ changing with distance (reaching $10\%$ at 1~AU) are displayed in Figure \ref{fig:PF_F_all} for four different values of $\psi$ within the range $0 \le \psi \le 0.8$.  The decreased presence of fluctuations near the Sun ($\psi > 0$) increases the amount of induced plasma waves which decreases the spectral index below the break energy, shown in Figure \ref{fig:SI_str_len_pert}.  For all simulated values of $\psi > 0$ no bump in the fluence spectra is present.  A reduced level of fluctuations near the Sun decreases positive density gradients which subsequently decreases the acceleration of electrons from re-absorption of plasma waves.  A smoother increase in spectral index below the break energy for increasing $\psi$ can thus be seen in Figure \ref{fig:SI_str_len_pert}

\section{Discussion and Conclusions}

The simulations show that fine structure of the background solar wind electron density caused plasma waves to be suppressed, with larger amplitudes and smaller length scales of density fluctuations having the largest effect.  This increased suppression for larger amplitudes is similarly observed for higher levels of turbulence ($\Delta n/n$) with Kolmogorov type density fluctuations.

For high levels of turbulence near the Sun, $\Delta n/n = 10\%$,  wave production by the electron beam became no longer sufficient for the generation of type III radio bursts.  It is possible to induce more plasma waves by increasing the density of the electron beam.  This solution requires at least two orders of magnitude more electrons, causing the beam to have around $1\%$ of the density of the background plasma.  Such high density electron beams become problematic when considering simultaneous HXR bursts assuming the upward electron beam has only $0.2\%$ of the downward electron beam density, found above 50~keV in \citet{Krucker_etal07}.

Increasing the level of plasma waves near the Sun without increasing beam density, the amplitude of density fluctuations can be reduced.  We implemented a radial dependence with closer distances to the Sun have a decreased turbulent intensity.  This is seen in observational scintillation techniques \citep{Woo_etal1995,Woo1996} and \emph{Helios} data \citep{MarschTu1990} in the fast solar wind.  The observed values for $\Delta n/n$ are as low as $0.3\%$ at distances $<0.1$~AU \citep{Woo_etal1995}.  A much higher magnitude of plasma wave energy density was achieved close to the Sun with smaller levels of fluctuations.

\begin{figure} \center
\includegraphics[width=0.49\columnwidth]{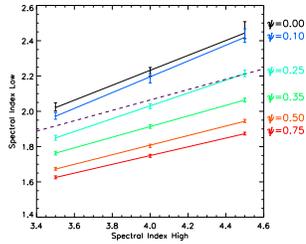}
\caption{Comparison between the high and low spectral index of fluence spectra of electrons near the Earth.  The dashed purple line is the best fit to the observational data of peak flux spectral indices \citep{Krucker_etal09}.}
\label{fig:SI_high_low_vary} \end{figure}

To estimate how density flucutations might radially evolve, we varied the initial conditions of the simulations.  We used a variety of different initial electron beam spectral indices ($\delta$ in Equation \ref{init_f}) and different radial dependence of density fluctuations ($\psi$ in Equation \ref{pertm3}).  Using the resulting fluence spectra near the Earth for each simulation, we compared the spectral indices above and below the break energy (Figure \ref{fig:SI_high_low_vary}).  The spectral index becomes smaller below the break energy for larger values of $\psi$.  We have also overplotted the correlation of spectral indices above and below the break energy of peak flux measurements taken from a statistical survey \citep{Krucker_etal09} of impulsive electron events detected by the three-dimensional Plasma and Energetic Particles experiment on the WIND spacecraft \citep{Lin_etal95}.  A level of fluctuations with $\psi$ around 0.25 would give a similar correlation to the observational line.   We note, however, that the observational line presented from \citet{Krucker_etal09} fitted a large scatter of data points.  The ratio of low:high spectral index for all simulated results presented in figure \ref{fig:SI_high_low_vary} lies between 0.42 and 0.58 which is within the narrow range presented in \citet{Krucker_etal09}.

A variety of simulation variables can affect the energy of the spectral break at the Earth:  the model of radial background density decrease, the density fluctuations, the initial spectral index of the beam, the number density of injected electrons,  the spatial distribution of injected electrons, the temporal nature of the injection, and the initial coronal background density where the electrons are injected.  The spectral index below the break energy of the resultant double power-law in fluence spectra near the Earth is increased when density fluctuations have a larger effect on the level of induced plasma waves.  It is important to note, however, the spectra below the break energy is only approximately a power-law.  The presence of density fluctuations causes fine structure to be present.  A bump around 10-20~keV was found, caused by acceleration of electrons at the back of the beam through absorbed plasma waves.  The onset of this bump appears to be close to the Sun where plasma wave energy density is high.  The magnitude of this bump is reflected in the size of the spectral index error bars in Figure \ref{fig:SI_high_low_vary} with a larger bump corresponding to a larger error.  With the prospect of Solar Orbiter and Solar Probe Plus, it is very attractive to extend these studies further to understand the spectral evolution of the electron beam between the Sun and the Earth.

\acknowledgments
This work is partially supported by a STFC rolling grant and STFC Advanced Fellowship (EPK). Financial support by the Royal Society grant (RG090411), and by the European Commission through the SOLAIRE Network (MTRN-CT-2006-035484) is gratefully acknowledged. The overall effort has greatly benefited from support by a grant from the International Space Science Institute (ISSI) in Bern, Switzerland.

\bibliographystyle{apj} \bibliography{inhom_1au}

\clearpage

\end{document}